\documentclass[%
 reprint,
groupedaddress,
 amsmath,amssymb,
 aps,
 prl
]{revtex4-1}

\usepackage[english]{babel}
\usepackage[utf8]{inputenc}
\usepackage{amsmath}
\usepackage[colorinlistoftodos]{todonotes}
\usepackage{verbatim} 
\usepackage{placeins}

\newcommand{\ket}[1]{|#1\rangle}
\newcommand{\braket}[2]{\langle{#1}|{#2}\rangle}

\newcommand{\bra}[1]{\langle#1|}

\newcommand*\diff{\mathop{}\!\mathrm{d}}

\newcommand*\dv[2]{\frac{\diff #1}{\diff #2}}
\newcommand*\pdv[2]{\frac{\partial #1}{\partial #2}}

\def\eq{\begin{eqnarray}}
\def\en{\end{eqnarray}}

\def\beq{\begin{eqnarray}}
\def\een{\end{eqnarray}}
\def\bfig{\begin{figure}}
\def\efig{\end{figure}}

\newcommand{\ip}[2]{\langle #1 | #2 \rangle}

 \usepackage[normalem]{ulem}

 	\definecolor{forestgreen}{rgb}{0.13, 0.55, 0.13}

\begin{document}

\title{Distinguishability and many-particle interference}
\author{Adrian~J.~Menssen$^{1,*}$, Alex~E.~Jones$^{1,2,*}$, Benjamin~J.~Metcalf$^{1}$, Malte~C.~Tichy$^{3}$, Stefanie~Barz$^{1}$, W.~Steven~Kolthammer$^{1}$, Ian~A.~Walmsley$^{1}$}
\affiliation{$^{1}$~Clarendon Laboratory, Department of Physics, University of Oxford, OX1 3PU, United Kingdom,\\
$^{2}$~Blackett Laboratory, Imperial College London, SW7 2BW, United Kingdom,\\
$^{3}$~Department of Physics and Astronomy, University of Aarhus, DK-8000 Aarhus C, Denmark,\\
$^{*}$~These authors contributed equally to this work.}
\begin{abstract}
Quantum interference of two independent particles in pure quantum states is fully described by the particles' distinguishability: the closer the particles are to being identical, the higher the degree of quantum interference.
When more than two particles are involved, the situation becomes more complex and interference capability extends beyond pairwise distinguishability, taking on a surprisingly rich character. Here, we study many-particle interference using three photons. We show that the distinguishability between pairs of photons is not sufficient to fully describe the photons' behaviour in a scattering process, but that a collective phase, the \textit{triad} phase, plays a role.
We are able to explore the full parameter space of three-photon interference by generating heralded single photons and interfering them in a fibre tritter. Using multiple degrees of freedom---temporal delays and polarisation---we isolate three-photon interference from two-photon interference. Our experiment disproves the view that pairwise two-photon distinguishability uniquely determines the degree of non-classical many-particle interference.
\end{abstract}
\maketitle

%


The famous Hong-Ou-Mandel (HOM) experiment in 1987 provided the first important example of non-classical two-photon interference~\cite{hong1987}. Two independent photons impinging on a beam splitter exhibit bunching behaviour at the output ports that cannot be explained by a classical field model.
The degree of bunching depends on how similar the two photons are in all degrees of freedom, for example time, frequency, polarisation, and spatial mode. This interference effect lies at the heart of photonic quantum information~\cite{knill2001} and has also become the standard tool for testing photon sources.
Extending the study of interference to many particles is of interest from a fundamental as well as from a technological viewpoint~\cite{Spagnolo2013, Tillmann2015, Maehrlein2015}. 
The scattering of multiple photons in linear networks is related to solving problems in quantum information processing, metrology, and quantum state engineering~\cite{Aaronson2011, Spagnolo2012, Broome2013, Spring2013b, Tillmann2013, Crespi2013, Graefe2014, Motes2015, Carolan2015}. Thus, understanding multiphoton interference is also of great relevance for practical applications.

Here, we demonstrate how many-particle interference is fundamentally richer than two-particle interference~\cite{Tichy2014}. Two situations with the \textit{same} pairwise distinguishability can lead to a \textit{different} output distribution. This is due to a phase, the \textit{triad} phase, that occurs only when more than two photons interfere.

We use independent photons and a tritter, a three-port symmetric beam splitter, as our tools for investigating multi-particle interference. We isolate the triad phase for the first time by interfering three photons in a tritter and exploiting multiple degrees of freedom, here time and polarisation.
We show that interfering three identical photons and varying time delays between them, as demonstrated in previous experiments~\cite{Spagnolo2013,Spring2016}, is not sufficient to study three-photon interference in full generality~\cite{Tichy2015}.
Further, we demonstrate that pairwise distinguishability between photons alone is not sufficient to fully describe the tritter's output statistics~\cite{Weihs1996}.
Our experiment allows us to isolate and tune the three-photon interference term as distinct from two-photon interference. Our work thus challenges the usual view that a general theory of photon (in)distinguishability can be reduced to time-delays~\cite{deGuise2014, Tillmann2015}.

%
\subsection{Theory}
%
The inner scalar product of two pure states $\ket{\phi_i}$ and $\ket{\phi_j}$ is:
\begin{equation}
	\braket{\phi_i}{\phi_j}=r_{ij}e^{i\varphi_{ij}}, 
\end{equation}
where $r_{ij} \in (0,1)$ is the real modulus and $\varphi_{ij} \in (0,2\pi)$ is the argument. The modulus $r_{ij}$ can be interpreted as a measure of the distinguishability of two photons in states $\ket{\phi_i}$ and $\ket{\phi_j}$, and equals zero (one) for two orthogonal (identical) states~\cite{Jozsa2000}. The argument $\varphi_{ij}$ has, so far, received little attention due to its irrelevance in two-photon interference.

\begin{figure}
\centering
\includegraphics[width=0.49\textwidth]{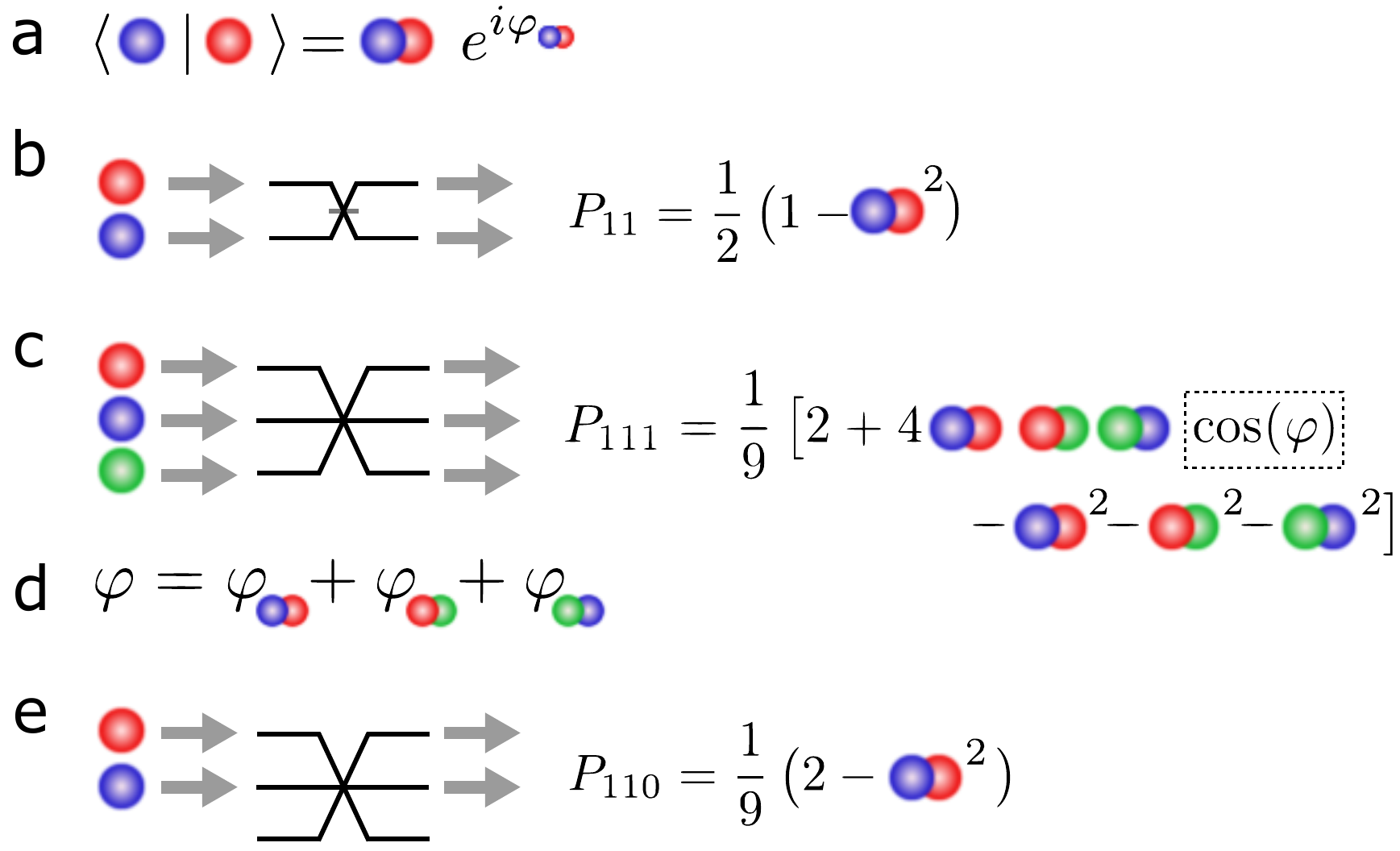}
\caption{\label{fig:Figure1} Interference of photons in balanced beam splitters and tritters. \textbf{a.}, \textbf{b.} The output statistics of two photon interfering in a beam splitter can be described via the pairwise distinguishability of the photons. \textbf{c.} In the case of a tritter, the output statistics depend on an additional phase $\varphi$. 
\textbf{d.} This \textit{triad} phase $\varphi$ is defined by the arguments of the pairwise complex scalar products.
\textbf{e.} $\varphi$ only occurs in the interference of more than two photons.
}
\end{figure}
We consider two examples of devices that can be used to probe interference: a beam splitter and a tritter.
The simplest device to probe interference is a balanced two-port beam splitter (see Fig.~\ref{fig:Figure1}a). When two photons $\ket{\phi_1}$ and $\ket{\phi_2}$ are injected into the beam splitter, the output statistics depend on the pairwise distinguishability of the incident photons: 
\begin{equation}\label{HOMcoinc}
P_{11}=\frac{1}{2}\left(1-r_{12}^2\right), 
\end{equation}
where $P_{11}$ is the probability for detecting one photon in each of the output ports.
If the photons are completely indistinguishable they always exit the same output port, in contrast to the classical behaviour.

A tritter maps three spatial input modes onto three spatial output modes (see Fig.~\ref{fig:Figure1}b); a linear transformation corresponding to a balanced tritter is given by the unitary matrix:
\begin{equation}\label{tritter}
U_{tritter}=\frac{1}{\sqrt{3}}\left(\begin{matrix}
1&1&1\\
1&\zeta^2&\zeta\\
1&\zeta&\zeta^2
\end{matrix}\right),
\end{equation}
where each output is equally likely and $\zeta=e^{i2\pi/3}$.

If we inject three photons into the tritter---a single photon in state $\ket{\phi_i}$ into each mode $i$ for each $i=1,2,3$---the probability $P_{111}$ of having one photon in each of the output modes of the tritter is (see Appendix)~\cite{Tichy2015a,Shchesnovich2015a}:
\begin{align}\label{eqn:P111}
P_{111} = \frac{1}{9}\left[2+ 4\:r_{12} r_{23} r_{31}\cos(\varphi)
-r_{12}^2-r_{23}^2-r_{31}^2\right]
\end{align} 
where we define the collective \textit{triad} phase $\varphi=\varphi_{12}+\varphi_{23}+\varphi_{31}$ as the sum of the three arguments.
The dependence on $\varphi$ appears only if the photons are partially distinguishable. 
If the states are orthogonal, the three moduli are zero; if they are identical, their scalar product will be equal to one and $\varphi$ vanishes. Similar expressions can also be derived for the probabilities of having two or three photons in one of the output modes of the tritter (see Appendix).

Note that a global phase applied onto one of the input states does not lead to any change in the triad phase~$\varphi$. 
Each phase $\varphi_{ij}$ is only defined up to a global arbitrary phase. The sum of the phases, the triad phase, has physical meaning and is a measurable quantity.
It remains unaffected by any global phase transformation and is crucial for the description of \textit{partially distinguishable} photons~\cite{Bergou2012, Sugimoto2010}.

However, dependence on the triad phase $\varphi$ only occurs in measurements with more than two photons. 
The two-photon output coincidence probabilities $P_{011}$ (one photon in outputs~2 and~3), $P_{101}$, $P_{110}$ when sending two photons into different input ports of the tritter (as in Fig.~\ref{fig:Figure1}e) are:
\begin{equation}\label{eqn:P011}
	P_{011}=P_{101}=P_{110}=\frac{1}{9}\left(2-r_{ij}^2\right),\mbox{ $i,j=1,2,3$; $i\neq j$}
\end{equation}
and depend only on the mutual distinguishability of the incident photons~$\ket{\phi_i}$ and $\ket{\phi_j}$ . 

\begin{figure*}
\centering
\includegraphics[width=0.8\textwidth]{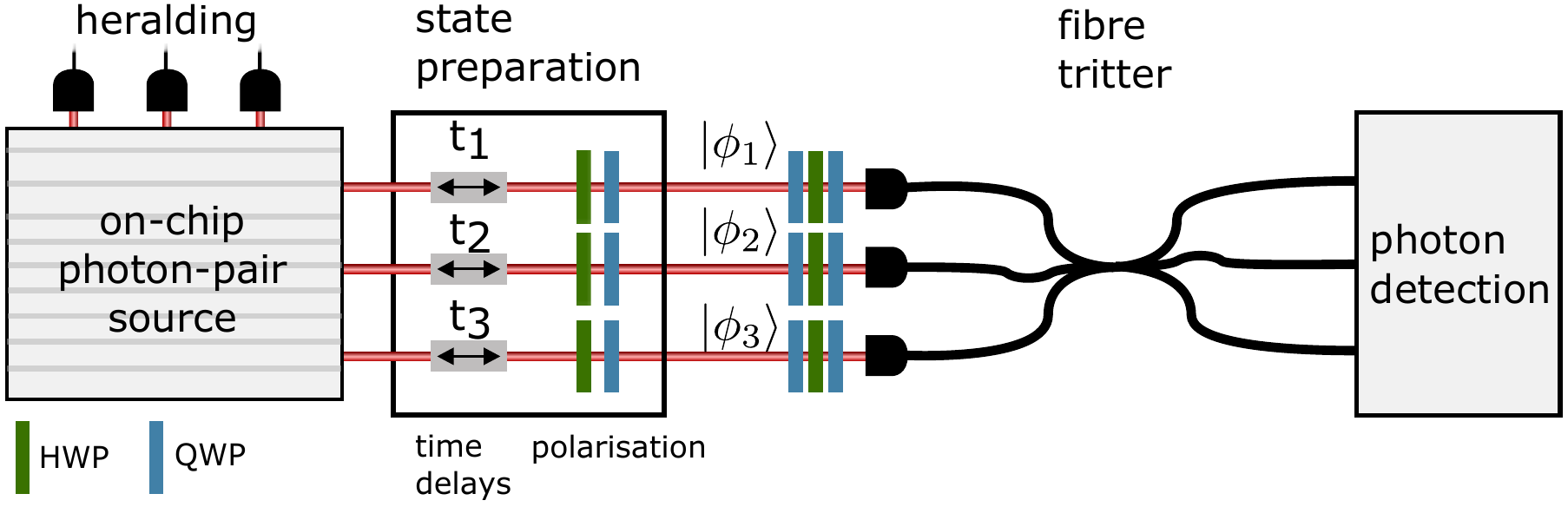}
\caption{\label{fig:Figure2} Scheme of the experimental setup.
 We generate three photon pairs using spontaneous four-wave mixing (SFWM) in silica-on-silicon waveguides~\cite{Spring2016}. 
Three signal photons~(at $680\,$nm) are used to herald three idler photons~(at $817\,$nm).
The relative temporal delays of the three photons are adjusted using delay stages.
We use sets of quarter-wave plates (QWPs) and half-wave plates (HWPs) to prepare the polarisation state of each photon and to compensate for polarisation rotations in the fibres.  The outputs of the fibre tritter are monitored using commercial avalanche photodiodes. As shown in the Appendix, we used additonal fibre beam splitters and a tritter for pseudo-number resolution.
The average rate of sixfold coincidences is about $0.5\,$Hz (see Appendix for further experimental details).}
\end{figure*}

%
\subsection{Probing the triad phase and genuine three-photon interference}
%

We introduce a convenient implementation that allows us to control the moduli $r_{ij}$ and the arguments $\varphi_{ij}$ independently.
We use two degrees of freedom for each spatial mode---time and polarisation---to show that the addition of non-identical polarisation states can be used to create a non-zero $\varphi$. We consider the following input states to the tritter (see Fig.~\ref{fig:Figure2}):
\begin{equation}
\ket{ \phi_i} = \ket{ t_i} \otimes \left( \cos \alpha_i \ket{H} + e^{i \eta_i} \sin \alpha_i \ket{V} \right)  \label{genpolstates} \\
\end{equation}
where $\ket{ t_i}$ is a temporal mode delayed by time $t_i$, $\ket{H}$ and $\ket{V}$ denote horizontal and vertical polarisation, respectively, and $i=1,2,3$ denotes the spatial mode.
Using only temporal modes, $\ket{t_1}, \ket{t_2}, \ket{t_3}$, and otherwise identical photons with symmetric spectral intensities, the triad phase would always vanish, since $\braket{t_1}{t_2}\braket{t_2}{t_3}\braket{t_3}{t_1}$ is real and non-negative (see Appendix for more information on temporal modes).

In a first experiment, we aim to probe the triad phase directly. As a first step, we prepare the photons with the same polarisation $\ket{H}$ in states
\begin{eqnarray}\label{eqn:staticpol1}
\ket{\phi_{i}} &=&\ket{ t_i}\otimes\ket{H} 
\end{eqnarray}
for $i=1,2,3$, which sets $\varphi=0$. 

\begin{figure*}
\centering
\includegraphics[width=1\textwidth]{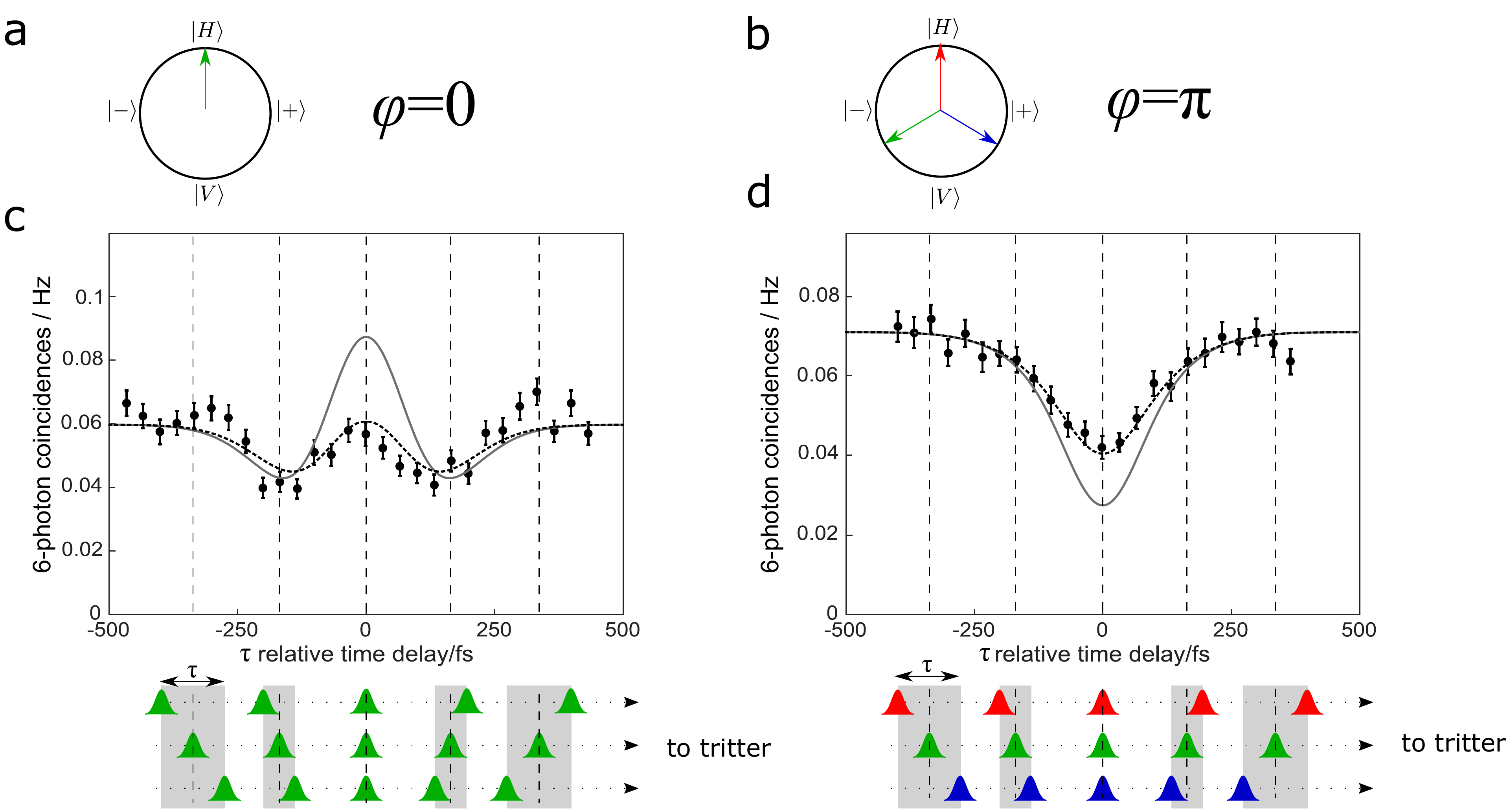}
\caption{\label{fig:Figure3} Experimental heralded three-photon coincidences at the output of a fibre tritter for two values of the triad phase $\varphi$.
\textbf{a}, \textbf{b}, We choose two polarisation configurations so that $\varphi=0$ (\textbf{a}) and $\varphi=\pi$ (\textbf{b}), see Eqns.~\ref{eqn:staticpol1} and~\ref{eqn:staticpol2}.
\textbf{c}, \textbf{d}, We measure heralded three-fold coincidences ($\propto P_{111}$) between the different output ports of the tritter whilst varying the temporal delays of the photons. As shown pictorially beneath the plots, we start in a configuration where the photons are completely distinguishable in time; two of the photons are then scanned symmetrically across the third photon ($t_1=t_2-\tau/2$, $t_3=t_2+\tau/2$). The grey boxes show the region of temporal overlap of the photons. The non-monotonic behaviour in \textbf{c} arises because $\varphi=0$ causes the three-photon interference term in Eqn.~\ref{eqn:P111} to have a contribution of opposite sign to those of the two-photon terms described by $r_{ij}^2$. In \textbf{d} $\varphi=\pi$ and so the contribution is of the same sign, resulting in monotonic behaviour of the statistic. 
The grey curve shows the theoretical prediction and the dashed black curve is calculated using a model which includes experimental imperfections (see main text for details). 
The absolute number of counts per point were between 200 and 350  (250 and 450) for \textbf{a} (\textbf{b}). Error bars are calculated from repeated measurements.
}
\end{figure*}

In the next step, we prepare photons in states (as depicted in the inset in Fig.~\ref{fig:Figure3}b):
\begin{eqnarray}\label{eqn:staticpol2}
\ket{\phi_{1}'} &=&\ket{ t_1}\otimes\ket{H} \\\nonumber
\ket{\phi_{2}'}&=&\ket{ t_2}\otimes\frac{1}{2}(\ket{H} +\sqrt{3}\ket{V})\\\nonumber
\ket{\phi_{3}'} &=&\ket{ t_3}\otimes\frac{1}{2}(\ket{H} -\sqrt{3}\ket{V}).
\end{eqnarray}
Here the scalar products $\braket{\phi_{1}'}{\phi_{2}'}=1/2 \braket{t_1}{t_2}$ and $\braket{\phi_{3}'}{\phi_{1}'}=1/2 \braket{t_3}{t_1}$, but $\braket{\phi_{2}'}{\phi_{3}'}=-1/2 \braket{t_3}{t_1}$, setting $\varphi=\pi$. These two configurations demonstrate that using polarisation as an additional degree of freedom allows us to vary the triad phase $\varphi$ (see Appendix for more details).
%

%
In a second experiment, we \textit{isolate} three-photon interference from two-photon interference. We explicitly show that control of $\varphi$ allows manipulation of the three-photon term whilst leaving the two-photon interference terms constant.
To do so, we prepare the following as input states to the tritter:
\begin{align}\label{eqn:dynamicpol}
	\ket{\phi_1''}&=\ket{t_1}\otimes \left[ \cos{\left(2\theta\right)}\ket{H}+i\sin{\left(2\theta\right)}\ket{V}\right]\\\nonumber
	\ket{\phi_2''}&=\ket{t_2} \otimes \left[\frac{1}{2}(\sqrt{3}\ket{H}+\ket{V})\right]\\\nonumber
	\ket{\phi_3''}&=\ket{t_3} \otimes \left[\frac{1}{2}(\sqrt{3}\ket{H}-\ket{V})\right], 
\end{align}
where the state $\ket{\phi_1''}$ depends on a polarisation rotation with angle $\theta$ and the polarisations of $\ket{\phi_2''}$ and $\ket{\phi_3''}$ are kept constant.
With these states, we obtain the following moduli
\begin{align}
r_{12}&=|\braket{t_1}{t_2}|\times\frac{1}{2}\sqrt{2+\cos(4\theta)}\\
r_{31}&=|\braket{t_3}{t_1}|\times\frac{1}{2}\sqrt{2+\cos(4\theta)}\\
r_{23}&=|\braket{t_2}{t_3}|\times\frac{1}{2} 
\end{align}
and the triad phase
\begin{align}
\varphi&=2\,{\rm Arg}\left(\sqrt{3}\cos(2\theta)+i\sin(2\theta)\right).
\end{align}
The angle $\theta$ affects both the triad phase $\varphi$ and the moduli $r_{12}$, $r_{31}$; the temporal state $\ket{t_1}$ only affects $r_{12}$ and $r_{31}$, but not $\varphi$.
Combining control of both $\theta$ and $\ket{t_1}$ allows us to manipulate $\varphi$ whilst $r_{12}$ and $r_{31}$ remain unchanged.
For example, to keep $r_{12}=r_{23}=r_{31}=1/2$, $\ket{t_1}$ must be chosen such that
\begin{equation}\label{eq:timediff}
		|t_1-t_2|=|t_1-t_3|=\sigma\sqrt{2 \ln[2+\cos(4\theta)]}
\end{equation}
with $t_2=t_3$ and $\sigma^2$ being the variance in time of the Gaussian wave packet (see Appendix).

Hence we can prepare three photons in such a way that three-photon measurements such as $P_{111}$ change with $\varphi$, whereas the two-photon measurements ($P_{110}$, $P_{101}$, $P_{011}$) remain constant (see Eqns.~(\ref{eqn:P111}) and~(\ref{eqn:P011})).


\begin{figure*}
\centering
\includegraphics[width=0.9\textwidth]{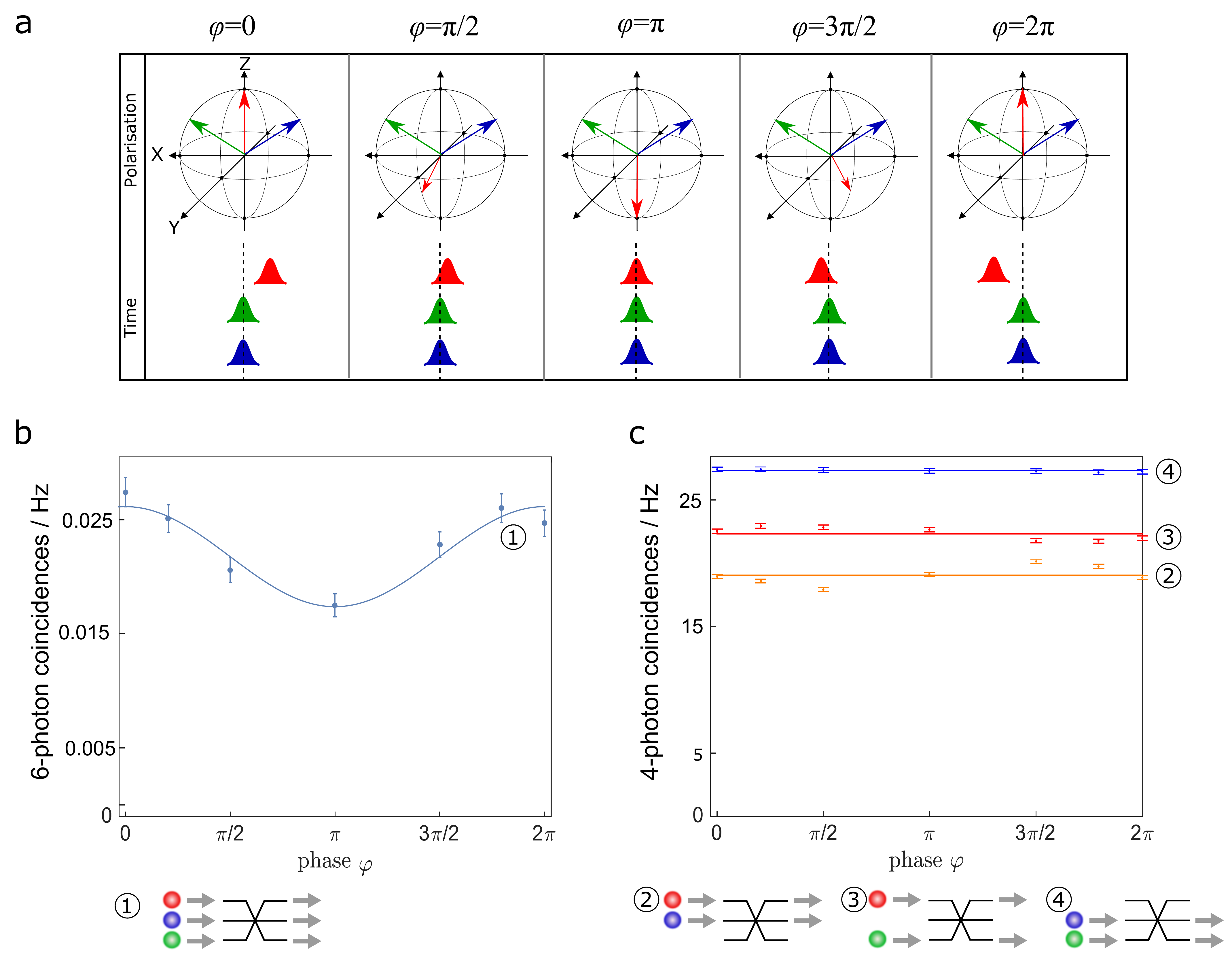}
\caption{\label{fig:Figure4} Isolating two-photon from three-photon interference. \textbf{a}, We vary the triad phase by rotating the polarisation of photon $\ket{\phi_1''}$ as given in Eqn.~\ref{eqn:dynamicpol}, leaving the polarisation states of the two other photons fixed. To keep the moduli $r_{12}$ and~$r_{31}$ constant, we adapt the temporal overlaps of the photons by tuning~$\ket{t_1}$.
\textbf{b}, The three-photon signal $P_{111}$ varies with the triad phase (absolute number of counts per data point is between~330 and~515). The plotted curve is a theory curve calculated based on our model of the experiment.
\textbf{c}, We plot a subset of two-photon distinguishability terms to demonstrate that these are kept constant. This experiment shows that the same pairwise distinguishability can lead to different degrees of multi-particle interference. 
}
\end{figure*}

%
\subsection{Experiment and Results}
%
To study the triad phase experimentally, we generate three heralded photons using spontaneous four-wave mixing (SFWM) in silica-on-silicon waveguides~\cite{Spring2016}. Using wave plates and delay stages, we prepare the polarisation and temporal state of each input photon before coupling into the fibres to the tritter (see Fig.~\ref{fig:Figure2} and Appendix for technical details). 

We first probe the triad phase $\varphi$ directly by choosing the input polarisations of the photons as given in Eqns.~(\ref{eqn:staticpol1}) and~(\ref{eqn:staticpol2}). 
By setting $t_1=t_2-\tau/2$, and $t_3=t_2+\tau/2$, and varying $\tau$ smoothly over the range shown in Fig.~\ref{fig:Figure3}c and~d, we tune the degree of two- and three-photon interference.
The results are shown in Fig.~\ref{fig:Figure3}; we see a clear qualitative difference in behaviour for the two cases of $\varphi=0$ and $\varphi=\pi$. In the former case we observe a W-like shape, whereas for the other case we observe a dip; deviations from the ideal curves are discussed below (see caption of Fig.~\ref{fig:Figure3}).

We then demonstrate genuine three-photon interference by choosing the input states as given in Eqn.~(\ref{eqn:dynamicpol}), but now setting the time delay differences as in Eqn.~\ref{eq:timediff}. 
We determined $\sigma$ from a set of two-photon HOM dips with polarisations chosen as in Fig.~\ref{fig:Figure4}a (first and third panel).
The results are shown in Fig.~\ref{fig:Figure4}; we observe good agreement of the measured curves with the theoretical prediction. The three-photon data follow a cosine shape as predicted by Eqn.~\ref{eqn:P111}.
The two-photon contributions $P_{110}$, $P_{101}$, $P_{011}$ (see Eqn.~\ref{eqn:P011}) are nearly constant and show fluctuations of only on average 6\%.
and the single photon detections at the tritter outputs vary only by a maximum of 3\% due to polarisation dependence.
This verifies that these two-photon contributions are independent of the arguments. 
Detailed analysis suggests that polarization dependence of the tritter contributes to these fluctuations (see Appendix).

Our experimental data, both in Fig.~\ref{fig:Figure3} and Fig.~\ref{fig:Figure4}, show the expected behaviour, but there are some deviations from the probabilities given by Eqns.~(\ref{eqn:staticpol1}),~(\ref{eqn:staticpol2}), and~(\ref{eqn:dynamicpol}).
These are primarily due to an imperfect tritter operation, imperfections in the photon preparation (polarisation, purity ($P>90\%$), distinguishability), and higher-order photon emission (squeezing parameter $\lambda=0.16$, see Appendix). Further, along with photons that are produced by the SFWM-process, uncorrelated photons are created in other processes such as Raman scattering and fluorescence~\cite{Spring2016}. To understand the influence of all these effects on the measured visibilities, we performed a simulation of our experiment.
Our model includes terms corresponding to up to $N=8$ photons in total (signals and heralds) and up to 3 uncorrelated noise photons. This provides sufficient accuracy as terms corresponding to higher photon numbers are negligibly small.
Based on our model, we calculated theory curves including realistic experimental parameters. These curves are shown in Fig.~\ref{fig:Figure3} and Fig.~\ref{fig:Figure4} as dashed lines and agree well with our measured data (see Appendix for a more detailed analysis) (see also~\footnote{We became aware that similar results to those in Fig.~\ref{fig:Figure4} have been obtained using a three-photon entangled state~\cite{Agne2016}}).

%
\subsection{Conclusion}


In this work, we identify and describe a new phase that arises at the level of three photons: the triad phase. This new phase manifests itself in quantum interference and therefore has implications for the scattering of many particles. In particular, the outcome of scattering events of more than two particles is determined not only by pairwise distinguishabilities of the particles' wavefunctions, but also on the collective properties of the particles.  
In this context, the triad phase initially emerges as a formal artifact~\cite{Tichy2011, Tan2013, Ra2013a, deGuise2014, Shchesnovich2013, Tamma2014, Tichy2014, Shchesnovich2015, Shchesnovich2015a, Tamma2015}; we show here that it is of physical relevance.
Two situations involving the scattering of multiple particles with the \textit{same} pairwise distinguishability can nevertheless exhibit a \textit{different} outcome for scattering depending on the triad phase. 

There is a formal similarity between the triad phase and the geometric phases that can be acquired by single photons, for instance in the Pancharatnam-Berry phase~\cite{Berry1987,Jozsa2000, Hartley2004, Kobayashi2010}. Scaling up our study to more than three photons is ongoing work, but for example four interfering photons can be described by six two-photon measurements and three three-photon measurements.

Our work has implications for both linear-optical quantum computing and general multiparticle scattering. It shows having truly indistinguishable particles is a crucial ingredient for all types of scattering experiments. However, our work also opens up new opportunities as the triad phase can be seen as a tool to engineer the output state of a scattering process.
Furthermore, extending applications such as boson sampling to partial distinguishabilities and using multiple degrees of freedom will be an interesting avenue to explore.


%


\section{Acknowledgements}
We thank Carlo Di Franco, Luca Rigovacca, Myungshik Kim, and Jan Sperling for helpful discussions.
A.J.M is supported by the James Buckee scholarship from Merton College.
A.E.J. is supported by the EPSRC Controlled Quantum Dynamics CDT.
M.C.T. acknowledges support from Danish Council for Independent Research and the Villum Foundation.
S.B. acknowledges support from the Marie Curie Actions within the Seventh Framework
Programme for Research of the European Commission, under the Initial Training Network PICQUE (Photonic Integrated Compound Quantum Encoding, grant agreement
no. 608062) and from the European Union’s Horizon 2020 Research and Innovation program under Marie Sklodowska-Curie Grant Agreement No. 658073. 
I.A.W. acknowledges an ERC Advanced Grant (MOQUACINO) and the UK EPSRC project EP/K034480/1.
\section{Author contributions}
A.J.M. and A.E.J. performed the experiments, theoretical modeling, and data analysis. B.J.M., S.B., and W.S.K assisted with data-taking and data analysis.
All authors discussed the results. M.C.T. developed the theory. B.J.M., S.B., W.S.K. and I.A.W conceived the project. S.B., W.S.K. and I.A.W. supervised the project.
All authors wrote the manuscript.

\newpage

\onecolumngrid
\appendix

\section{Appendix: Additional theory}

\subsection{Transition probability for three partially distinguishable bosons in a three mode-setup}
%

We inject three partially distinguishable bosons into the three input modes of a scattering setup described by a unitary matrix $U$. The distinguishing degrees of freedom of the bosons -- in our case the photon polarisation and the time-of-arrival -- are described by the states $\ket{\phi_1}, \ket{\phi_2}, \ket{\phi_3}$. The mutual pairwise distinguishabilities are then encoded in the positive semi-definite hermitian matrix $\mathcal{S}_{j,k} = \braket{\phi_j}{\phi_k} = r_{jk} e^{i \varphi_{jk}} $, which accommodates both the scalar product moduli $r_{jk}$ as well as the relative phases $\varphi_{jk}$. The probability to find one particle in each output mode is obtained as a sum over all possible double-sided Feynman diagrams~\cite{Tichy2015a}, giving a multidimensional permanent
\eq 
P_{111} = \sum_{\sigma, \rho \in S_3} \prod_{j=1}^3 U_{\sigma_j, j} U_{\rho_j, j}^\star \mathcal{S}_{\rho_j, \sigma_j} , \label{multidimperm}
\en
comprising fully distinguishable particles ($\mathcal{S}= \text{diag}(1,1,1)$, $r_{jk}=\delta_{j,k}$) and perfectly identical bosons ($\mathcal{S}_{j,k}=1$, $r_{jk}=1$) as extremal cases.

Our aim is to understand the dependence of three-photon interference on the distinguishability parameters $r_{jk}$ and $\varphi_{jk}$ in detail. For this purpose, we write out the sums over the permutation group $S_3$ explicitly, 
\eq 
P_{111} &=& \text{perm}(U*U^\star) +  r_{12}^2\, \text{perm}(U*U_{2,1,3}^\star) + r_{31}^2\, \text{perm}(U*U_{3,2,1}^\star)+r_{23}^2\, \text{perm}(U*U_{1,3,2}^\star) \nonumber \\
 && + 2 \Re\{ r_{12}r_{23}r_{31} e^{i(\varphi_{12}+\varphi_{23}+\varphi_{31})}\text{perm}(U*U_{2,3,1}^\star)\} , \label{generalexpr}
\en
where $U^\star_{x,y,z}$ is the element-wise complex conjugated matrix $U^\star$ with the \emph{rows} (corresponding to the input modes) permuted as $(1,2,3) \rightarrow (x,y,z)$, i.e.~for $(x,y,z)=(1,2,3)$, we leave the matrix unchanged, while for $(x,y,z)=(2,1,3)$, we exchange the first two rows. The product $X*Y$ is meant as Hadamard element-wise multiplication,  not the usual matrix-product. 

For final states with $s_j$ particles in the $j$th output mode, we adapt Eqn.~(\ref{generalexpr}) formally by replacing $U$ by a matrix of the same dimensions that repeats the $j$th column of $U$ $s_j$ times , i.e.~the column multiplicity reflects the final mode population. To ensure the proper normalization of the final result, a factor $(\prod_{j=1}^3 s_j!)^{-1}$ needs to be included, where $\vec s=(s_1,s_2,s_3)$ is the mode occupation list of the final state. 

We see clearly how the dependence on the scattering matrix $U$ is separated from the dependence on the scalar products $\mathcal{S}$. For a fixed scattering matrix $U$, the output signals depend on precisely six parameters, $r_{12}, r_{31}, r_{23}, \varphi_{12}, \varphi_{23}, \varphi_{31}$, of which only four have physical significance: the three scalar product moduli $r_{12}, r_{23}, r_{31}$ and the collective triad phase $\varphi = \varphi_{12}+\varphi_{23}+\varphi_{31}$. Whereas each relative phase $\varphi_{jk}$ can be transformed away by a global phase transformation, the  \emph{sum} of the three relative phases -- the collective triad phase $\varphi$ -- remains independent of any choice of basis or global phase. The dependence of scattering probabilities on a phase that describes the particles' collective indistinguishability has no precedent in single- or two-particle scattering. The triad phase only carries physical meaning in the context of the full three-particle state, and is thus a purely collective quantity.

\subsection{Event probabilities in the symmetric tritter}
For a symmetric tritter, 
\eq
U = \frac{1}{\sqrt 3} \left(\begin{array}{ccc} 
1 & 1 & 1 \\
1 & e^{i \frac{4 \pi}{3} } & e^{i \frac{2 \pi}{3} } \\
1 & e^{i \frac{2 \pi}{3} } & e^{i \frac{4 \pi}{3} }
\end{array} \right) ,
\en 
the output event probabilities take particularly simple forms: 
\eq\label{eqn:probs}
P_{111} &=& \frac{1}{9}\left[2+ 4\:r_{12} r_{23} r_{31}\cos(\varphi)-r_{12}^2-r_{23}^2-r_{31}^2\right] \\
P_{300} &=& P_{030} =  P_{003}=  \frac{1}{27} \left(1+r_{12}^2+r_{23}^2+r_{31}^2 + 2 r_{12} r_{23} r_{31} \cos (\varphi) \right),  \\
P_{120} &=& P_{012} = P_{201} = \frac{1}{9} \left(1-2 r_{12} r_{23} r_{31} \cos (\varphi+\pi/3)\right), \\
P_{021} &=& P_{210} = P_{102} = \frac{1}{9} \left(1-2 r_{12} r_{23} r_{31}   \cos (\varphi-\pi/3) \right).
\en 
Here, $P_{ijk}$ denotes the probability of measuring $i$ photons in output mode 1, $j$ photons in output mode 2, and $k$ photons in output mode 3.
The probability to find two particles in one output mode vanishes for indistinguishable photons, a result of the suppression law for Fourier matrices~\cite{Tichy2010}.

\section{Internal space dimensionality and triad phase}
 The triad phase $\varphi=\varphi_{12} + \varphi_{23} +\varphi_{31} $ encodes a purely collective property, which can only take non-trivial values for partially distinguishable particles: If all pairs of particles are mutually perfectly indistinguishable, we have $r_{ij}=1$, such that $\ket{\phi_1} \propto \ket{\phi_2} \propto \ket{\phi_3}$. The three states then span a trivial one-dimensional Hilbert-space, and $\varphi = 0$. On the other hand, when the particles are fully distinguishable, all scalar products $r_{12}$, $r_{23}$, $r_{31}$ vanish, and the value of the phase $\varphi$ is neither defined, nor does it have any impact on any observable, since it comes only in conjunction with the product $r_{12} r_{23} r_{31}$. In this case, the three states span a three-dimensional Hilbert-space. 
 
In the intermediate case in which all particles are neither mutually distinguishable nor indistinguishable ($0< r_{ij} < 1$ for all $i\neq j$), the question  arises whether the three internal states $\ket{\phi_1}, \ket{\phi_2}, \ket{\phi_3}$ span a three- or merely a two-dimensional Hilbert-space. This question arises, e.g. when three photons are deliberately prepared in different polarisation states, but are supposed to be indistinguishable in all other degrees of freedom. 

In order to see how the measurement of the triad phase $\varphi$ can resolve this question, let us first consider three states living in a qubit-like two-dimensional Hilbert-space. Without restrictions to generality, we can then find two states $\ket{0}$ and $\ket{1}$, such that
\eq
\ket{\phi_1} &=& \ket{0} \\
\ket{\phi_2} &=& \cos{\alpha} \ket{0} + \sin \alpha \ket{1} \\
\ket{\phi_3} &=& \cos \beta \ket{0}  + e^{i \gamma} \sin \beta \ket{1}  ,
\en
where $0 < \alpha, \beta < \pi/2$, $0 \le \gamma < 2 \pi$. We note that the three states are described by three parameters -- precisely those required to describe the relative positions of three points on a Bloch-sphere describing a qubit. In Eqn.~4 (main paper), however, four parameters -- $r_{12}, r_{23}, r_{31}, \varphi$ -- dictate the degree of three-particle interference.

By evaluating the relevant scalar products
\eq 
\braket{\phi_1}{\phi_2} &= & \cos \alpha \\
\braket{\phi_2}{\phi_3} &= & \cos \alpha \cos \beta + e^{-i \gamma} \sin \alpha \sin \beta\\
\braket{\phi_3}{\phi_1} &= & \cos \beta , 
\en
we can express $\gamma$ as a function of $r_{12}, r_{23}, r_{31}$, to see that  $\varphi$ is fixed by the three scalar product moduli $r_{12}, r_{23}, r_{31}$, i.e.~ \eq 
\varphi = \varphi_{2d}(r_{12}, r_{23}, r_{31}) 
\en In that sense, when restricted to a qubit-like Hilbert-space, the triad phase is not an independent degree of freedom, but it is fully fixed by the geometry of the three vectors on the Bloch sphere, or, equivalently, by $r_{12}, r_{23}, r_{31}$. 

When we lift the restriction to a qubit-like Hilbert-space and admit states of the more general form
\eq
\ket{\phi_1} &=& \ket{0} \\
\ket{\phi_2} &=& \cos{\alpha} \ket{0} + \sin \alpha \ket{1} \\
\ket{\phi_3} &=& \cos \epsilon \cos \beta \ket{0}  + \cos \epsilon e^{i \gamma} \sin \beta \ket{1}  + \sin \epsilon \ket{2} ,
\en
the new parameter $\epsilon$ describes to what extent the third state $\ket{\phi_3}$ lives outside the Hilbert-space spanned by $\ket{\phi_1}$ and $\ket{\phi_2}$. Hence, we now need four parameters to describe the three states, and even when $r_{12}, r_{23}, r_{31}$ are fixed, $\varphi$ now remains an independently tuneable parameter. 

As a consequence, the three-photon measurements yielding $r_{12}, r_{23}, r_{31}$ and $\varphi$ reveal whether or not the three states $\ket{\phi_1}, \ket{\phi_2}, \ket{\phi_3}$ span a three-dimensional Hilbert-space ($\epsilon >0$) or merely a two-dimensional one ($\epsilon=0$). The latter is clearly a collective property of all three states, and invisible to any combination of two-photon measurement data, which only yield $r_{12}, r_{23}, r_{31}$, but not the triad phase $\varphi$. In an experiment, a measurement of $\varphi$ that is compatible to  $\varphi_{2d}(r_{12}, r_{23}, r_{31}) $ implies that the three photons live in a two-dimensional space, while a measurement $\varphi$ that is incompatible with  $ \varphi_{2d}(r_{12}, r_{23}, r_{31}) $ shows that the three photons are distinguishable in more than a qubit-like degree of freedom.

\section{Dependence of triad phase on delay in three-photon interference}
\paragraph{} A single photon in spectral mode $\Psi$ is denoted
	\begin{equation} \label{eq:AD_BBMode}
	\ket{\Psi} = \int \diff \omega \tilde \psi(\omega) \ket{\omega}
	\end{equation}
where the state of a single photon with angular frequency $\omega$ is given by $\ket{\omega}$ and the mode is described by the complex-valued spectral amplitude $\tilde\psi(\omega)$. The same state can be described in the temporal domain with the mode transformation $\ket{\omega}=(2\pi)^{-1/2}\int \diff \tau \exp(i \omega \tau) \ket{\tau}$, where $\ket{\tau}$ describes a single photon with arrival time $\tau$. With this substitution, we find $\ket{\Psi} = \int \diff \tau \psi(\tau) \ket{\tau}$ where the temporal amplitude $\psi(\tau)$ is the inverse Fourier transform of $\tilde \psi(\omega)$.

If we delay a single photon initially in mode $\Psi$ by a time $t$, the resulting state is
	\begin{equation} \label{eq:AD_delayedMode}
	\ket{t}_\psi\equiv \int \diff \tau \psi(\tau-t) \ket{\tau}= \int \diff \omega e^{-i t\omega} \tilde \psi(\omega) \ket{\omega}.
	\end{equation}
In the following, we write this as $\ket{t}$ and forgo the subscript since only one initial mode will be considered. The inner product of two single-photons in initial mode $\Psi$ delayed by times $t_1$ and $t_2$ is
	\begin{equation} \label{eq:AD_innerProduct}
	\ip{t_1}{t_2} = \int \diff \omega e^{-i(t_2-t_1)\omega} |\tilde \psi(\omega)|^2 = \mathcal{F}\left[ |\tilde\psi|^2\right](t_2-t_1) \equiv \zeta(t_{21})
	\end{equation}
where the function $\zeta$ is defined as the Fourier transform of the spectral intensity $|\tilde \psi(\omega)|^2$ and $t_{ij} = t_i-t_j$.

Now consider the triad phase of three photons each in initial mode $\Psi$ with a distinct delay
	\begin{equation} \label{eq:AD_triad}
	\varphi = \arg \left[ \ip{t_1}{t_2}\ip{t_2}{t_3}\ip{t_3}{t_1} \right] = \arg \left[ \ip{t_1}{t_2}\right]+\arg \left[\ip{t_2}{t_3}\right]+\arg \left[\ip{t_3}{t_1}\right].
	\end{equation}
We are interested in the conditions for which $\varphi$ is independent of the delays. Accordingly we require that derivative with respect to delay vanishes 
	\begin{align} 
	\dv{\varphi}{t_1} &= \pdv{\varphi}{(t_{21})}\dv{(t_{21})}{t_1} + \pdv{\varphi}{(t_{13})}\dv{(t_{13})}{t_1}\\ \nonumber
			&= \dv{}{s}\big[ \arg\zeta(s) \big]\bigg|_{t_{13}} - \dv{}{s}\big[ \arg\zeta(s) \big]\bigg|_{t_{21}} = 0. 
	\end{align}
For this to be true for all values of $t_2$ and $t_3$, it must be that $\dv{}{s}\left[ \arg \zeta(s)\right]$ is independent of $s$. Therefore $\arg \zeta(s)$ is linear
	\begin{equation} \label{eq:AD_line}
	\arg\zeta(s) = \theta_0 + \Omega s.
	\end{equation}
The inverse Fourier transform of $\zeta$, the spectral intensity, is real. It follows that $\zeta(-s)=\zeta(s)^*$. Therefore, $\arg \zeta(s)$ is odd and $\theta_0=0$.

If $\varphi$ is independent of the delays, we can therefore write
	\begin{equation} \label{eq:AD_result}
	\zeta(s) = |\zeta(s)| \mathrm{e}^{i\Omega s} = \mathcal{F}\left[ |\tilde\psi(\omega)|^2\right](s).
	\end{equation}
Consider spectral intensity functions for which $\Omega=0$, which we denote as $|\tilde\psi_0(\omega)|^2$. For this case, since $|\tilde\psi_0(\omega)|^2$ and its Fourier transform $\zeta$ are real valued, $|\tilde\psi_0(\omega)|^2$ is an even function. Now consider the general case $|\tilde\psi_\Omega(\omega)|^2$ for non-zero $\Omega$. The shift property of the Fourier transform along with Eqn.\, (\ref{eq:AD_result}) tells us that $|\tilde\psi_\Omega(\omega-\Omega)|^2$ is an even function. Therefore we conclude that the triad phase $\varphi$ is independent of delays if the three photons start in identical spectral modes for which the spectral intensity is symmetric about its mean value.

\section{Inner products of Gaussian wavepackets with relative delays}

The state of a single photon in the time-frequency modes $(\tau,\omega)$, delayed by time $t$ is given by:
\begin{equation}
\ket{t}=\int d\tau\phi(\tau-t)\ket{\tau}
\end{equation}
For a Gaussian wave-packet delayed by time t, central frequency $\Omega$ and variance in time $\sigma^2$,
$\phi(\tau;t)$  takes the form:

\begin{equation}
\phi(\tau-t)=\Big( \frac{1}{\pi\sigma^2}\Big)^{1/4}e^{-\frac{(\tau-t)^2}{2\sigma^2}+i\Omega(t-\tau)}
\end{equation}

We can express the overlap of the temporal modes of two photons with identical Gaussian spectra at times $t_1$ and $t_2$ as:
\begin{equation}
\braket{t_1}{t_2}=\int_{-\infty}^{\infty}\phi^*(\tau-t_1)\phi(\tau-t_2)d\tau=e^{-\frac{(t_1-t_2)^2}{4\sigma^2}-i\Omega(t_1-t_2)}
\end{equation}
Next, we show that for time-delays the products of overlaps that appear in the expressions for the multi-photon coincidence probabilities are always real and positive and hence do not give rise to a triad phase.
In the case of the two photon interference terms, which contain expressions of the form 
\begin{equation}
\braket{t_1}{t_2}\braket{t_2}{t_1}=|\braket{t_1}{t_2}|^2
\end{equation}
this is easy to see, as the expression is purely real. For the three photon interference term:
\begin{equation}
\braket{t_1}{t_2}\braket{t_2}{t_3}\braket{t_3}{t_1}=e^{-((t_1-t_2)^2+(t_2-t_3)^2+(t_3-t_1)^2)/(4\sigma^2)-i\Omega(t_1-t_2+t_2-t_3+t_3-t_1)}=e^{-((t_1-t_2)^2+(t_2-t_3)^2+(t_3-t_1)^2)/(4\sigma^2)}
\end{equation}
As we can see this expression is also real. This also holds for any number of photons. 

%
\section{Experimental details}
\begin{figure}
\centering
\includegraphics[width=0.9\textwidth]{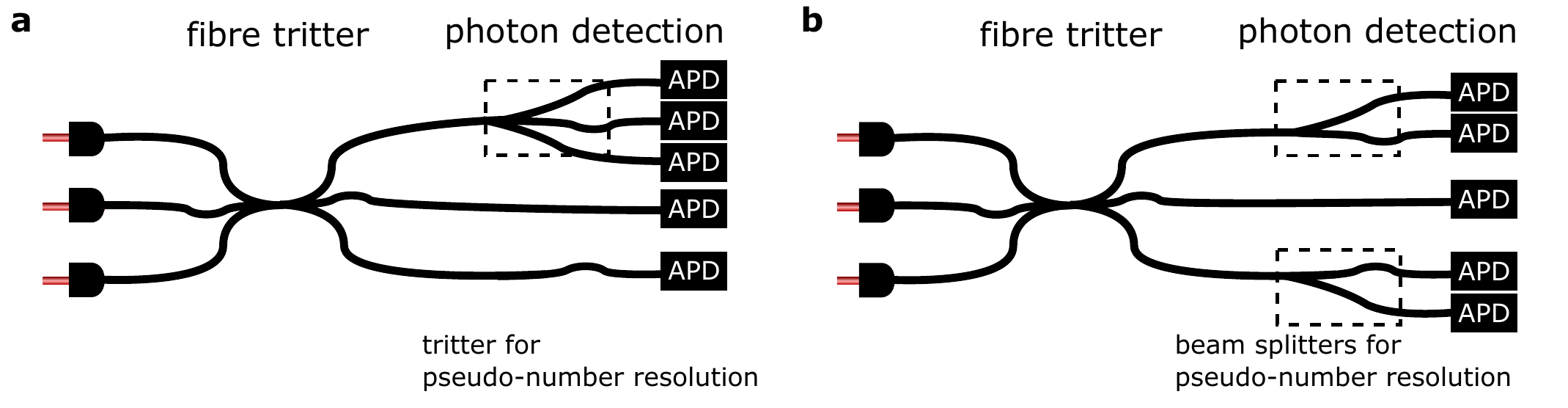}
\caption{\label{fig:pseudonumberres}
In our experiment we use additional fibre beam splitters and tritters to allow pseudo-number resolution. 
\textbf{a. }For monitoring genuine three-photon interference, the first and third outputs are connected to fibre beam splitters.
\textbf{b. }For probing the triad phase, we used a single tritter on the first output mode of the interference tritter.}
\end{figure}

%
We pump three separate waveguides in a silica-on-silicon chip~\cite{Spring2016} with a Ti-Saph femtosecond pulsed laser running at
$80\,$MHz and $740\,$nm ($130\,$fs pulses).
In each guide, we generate one signal and one idler photon at $680\,$nm and $817\,$nm, respectively. 
The pump has an orthogonal polarisation to the daughter photons and so
can be separated using a polarising beam splitter after the chip. The signal and idler
are spatially separated using a dichroic mirror before final filtering
to remove residual pump and to factor out their spectral components,
removing spectral correlations to give pure single photons.
Pumping three of these guides yields three signal photons and three
idler photons. By heralding on the former using three silicon APDs, we are
left with three heralded, highly pure identical single photons with
central wavelength $817\,$nm at a rate of about $0.5\,$Hz when pumping with
$130\,$mW per guide. The setup for measuring different output configurations after the interference tritter is shown in Fig.~\ref{fig:pseudonumberres}.

\section{Raw experimental data}

\subsection{Polarisations set for $\varphi=0$ (cf. Equation 7 in main paper)}
\subsubsection{HOM dips for temporal alignment of photons}
In order to align the generated photons temporally and verify their indistinguishability, we perform heralded HOM measurements for the three possible pairs injected into the tritter. We also use these to verify our polarisation state preparations. The results are shown in the Figures below.

\begin{figure}[h!]
  \centering
  \includegraphics[width=0.6\linewidth]{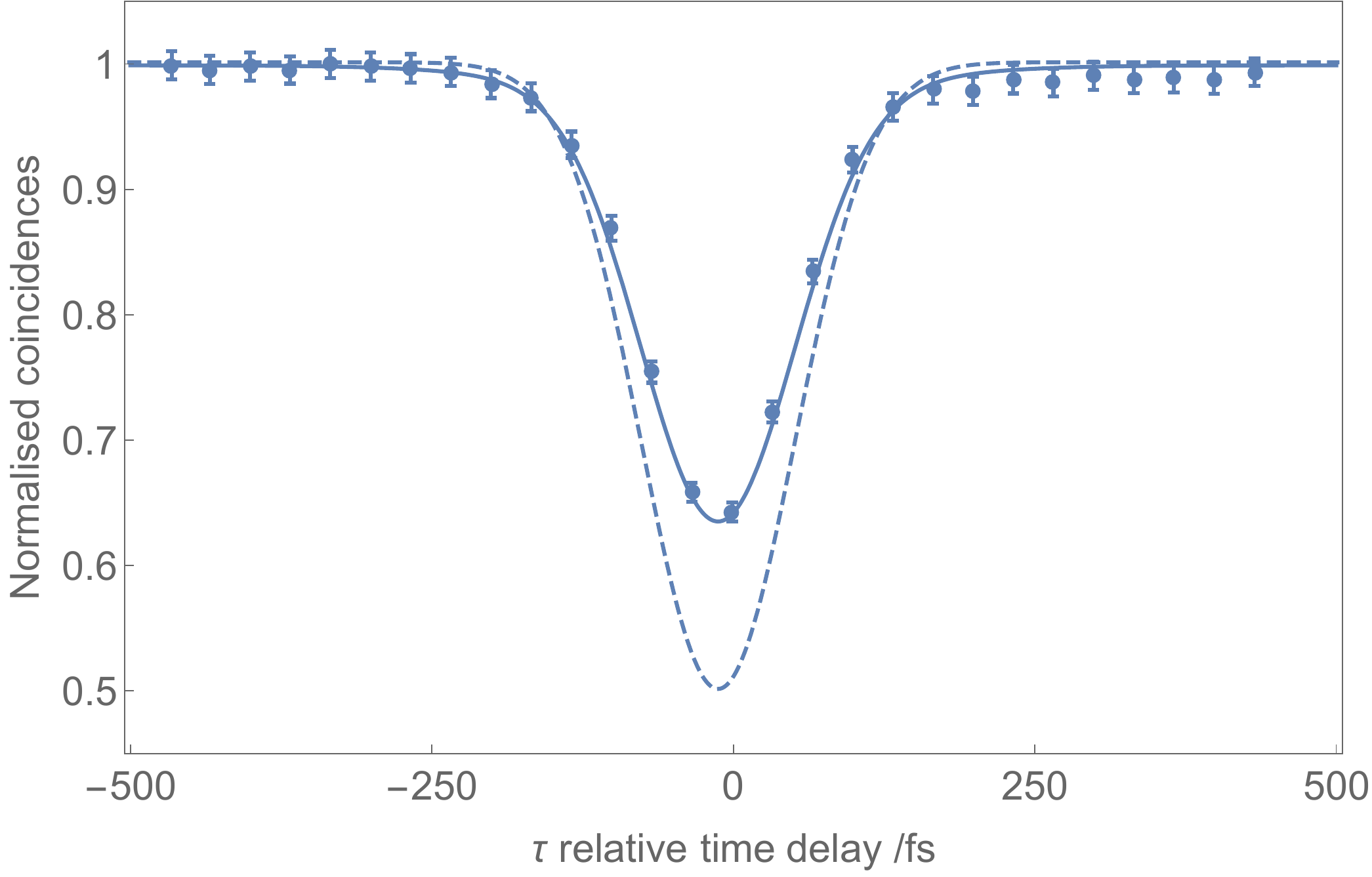}
  \caption{Plot of normalised heralded two-photon coincidences through the tritter when the injected photons have identical polarisations. In this case we inject photons into the first and second tritter inputs and monitor the first and second output ports. The solid line is the fit using our simulation, whilst the dashed line is an ideal theory curve.}
  \label{fig:HOM1212aligned}
  \end{figure}

For the case where the photons have identical polarisation, we expect a theoretical visibility of 50\% (since the two-photon coincidence probability is $P_{110}=\frac{1}{9}\times(2-|\braket{\phi_{1}}{\phi_{2}}|^{2}$) for a tritter, and so the visibility should be half the scalar product magnitude), we record closer to 40\% due to all effects mentioned in the main paper.  The dip in Figure~\ref{fig:HOM1212aligned} is twice as narrow as the others, corresponding to the dip between two photons which are both being translated in time on injection. The other two dips are from when only one of the photons injected into the tritter is translated in time (see Figure 3 in main text). The dips are all centred such that the three photons overlap in time when the stages are at their zero positions.

\begin{figure}[h!]
  \centering
  \includegraphics[width=0.6\linewidth]{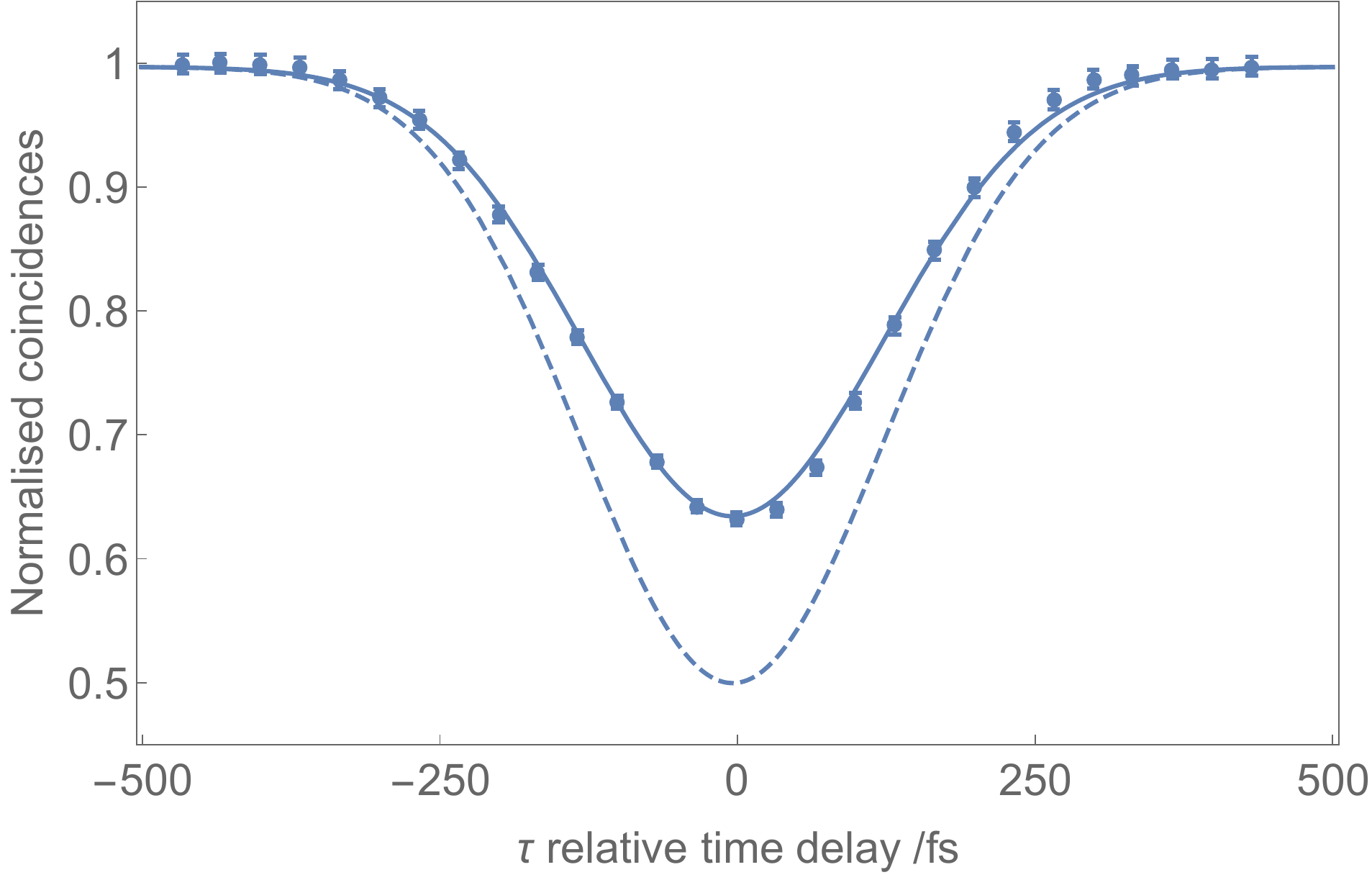}
  \caption{Plot of normalised heralded two-photon coincidences through the tritter when the injected photons have identical polarisations. In this case we inject photons into the first and third tritter inputs and monitor the first and third output ports. The solid line is the fit using our simulation, whilst the dashed line is an ideal theory curve.}
  \label{fig:HOM1313aligned}
\end{figure}
\begin{figure}[h!]
  \centering
  \includegraphics[width=0.6\linewidth]{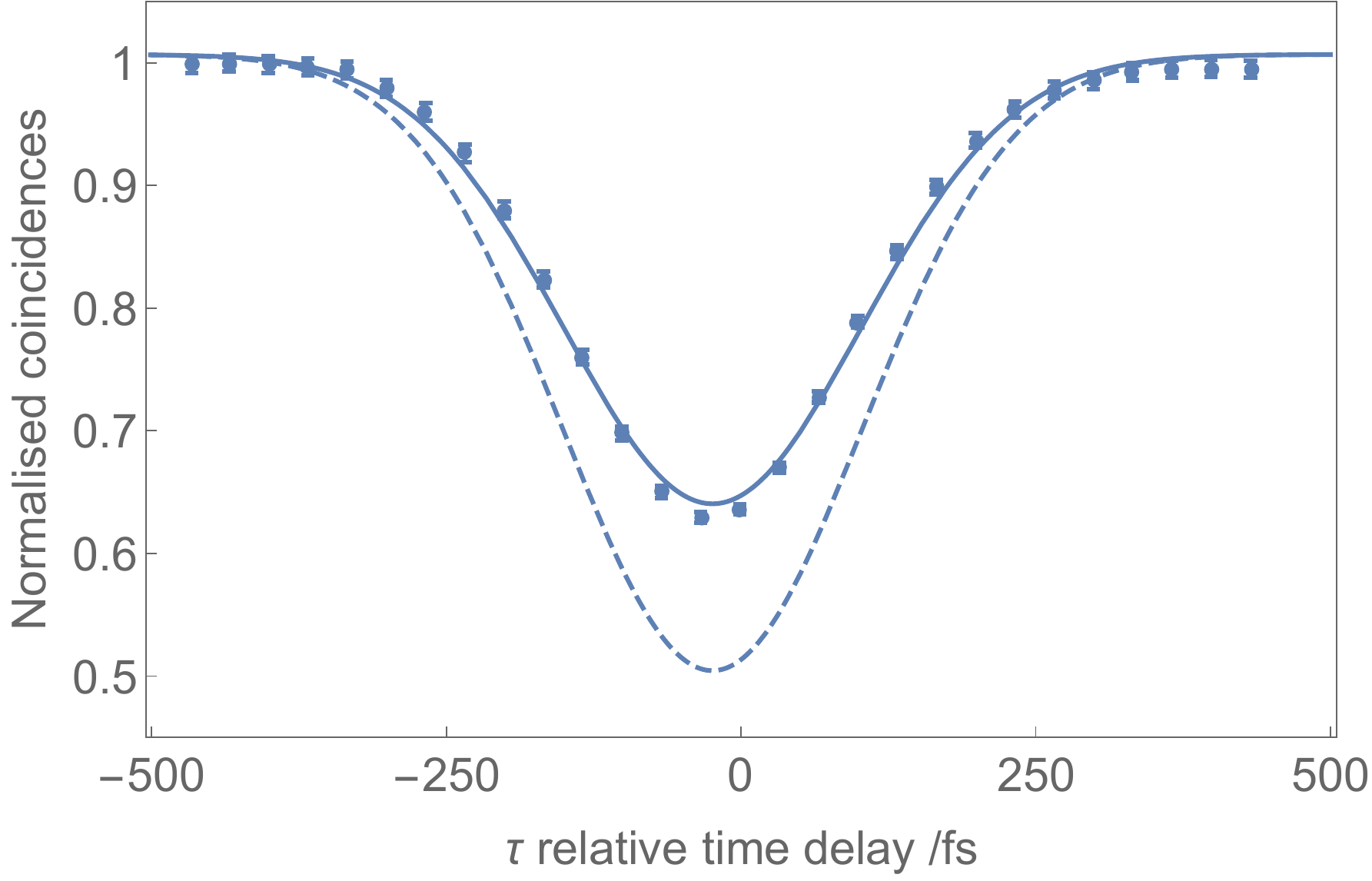}
  \caption{Plot of normalised heralded two-photon coincidences through the tritter when the injected photons have identical polarisations. In this case we inject photons into the second and third tritter inputs and monitor the second and third output ports. The solid line is the fit using our simulation, whilst the dashed line is an ideal theory curve.}
  \label{fig:HOM2323aligned}
\end{figure}

\clearpage
\subsubsection{Additional output event plots}
Here we present plots for count rates corresponding to $P_{210},P_{201},P_{300}$ in the case where all photons are injected into the tritter with the same polarisation. In the ideal case when all photons are completely indistinguishable in time and polarisation, $P_{210}=P_{201}=0$ and these outputs are completely suppressed~\cite{Tichy2010}. Our simulations demonstrate this is not the case when taking into account experimental imperfections, and the visibility is reduced from 100\% to around 57\%. The theory and simulation curves have been rescaled for comparison with experimental count rates.
\begin{figure}[h!]
  \centering
  \includegraphics[width=0.6\linewidth]{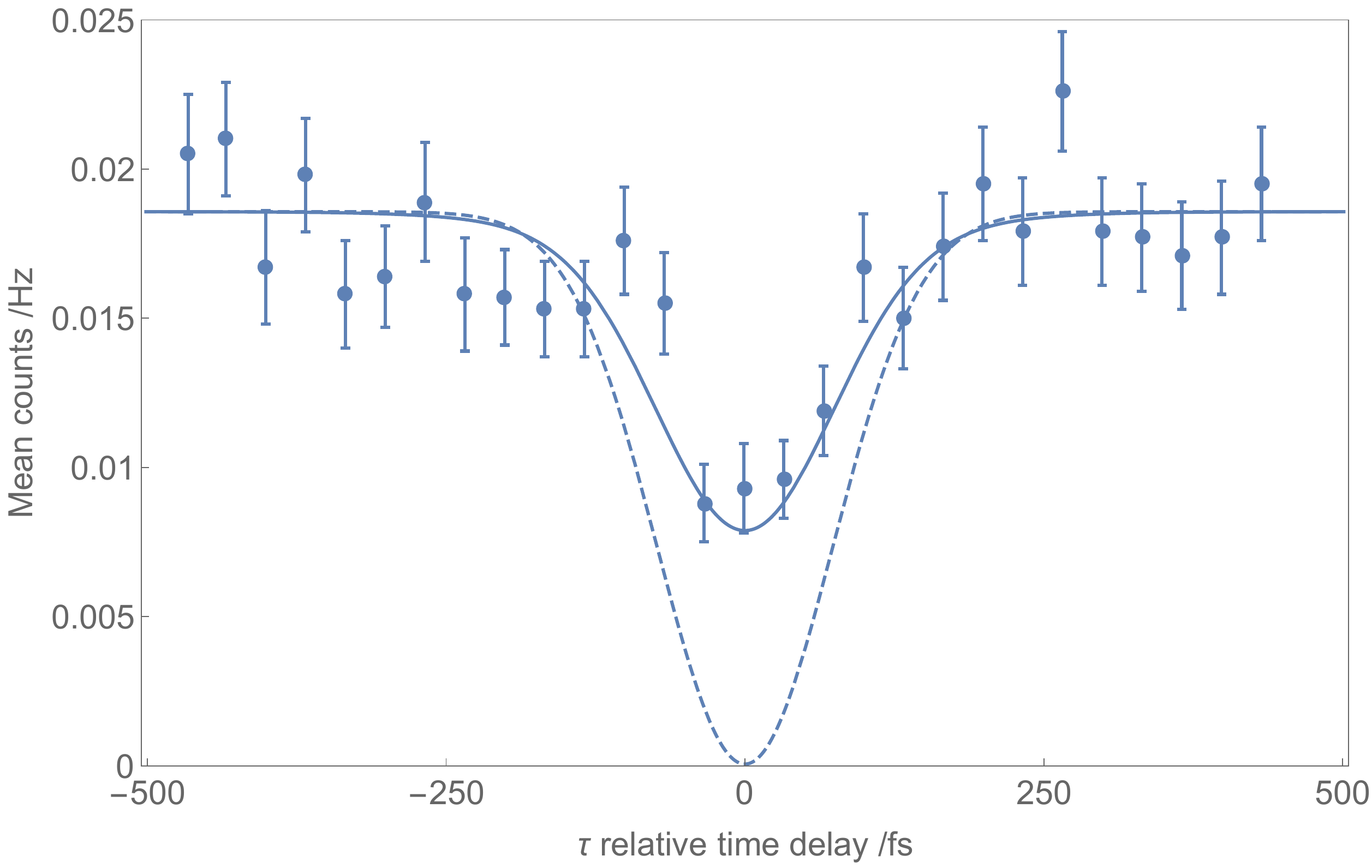}
  \caption{Plot of mean counts for the outputs corresponding to $P_{210}$ as illustrated in Figure~\ref{fig:pseudonumberres}a. The solid line is the fit using our simulation, whilst the dashed line is an ideal theory curve.}
  \label{fig:210aligned}
\end{figure}
\begin{figure}
  \centering
  \includegraphics[width=0.6\linewidth]{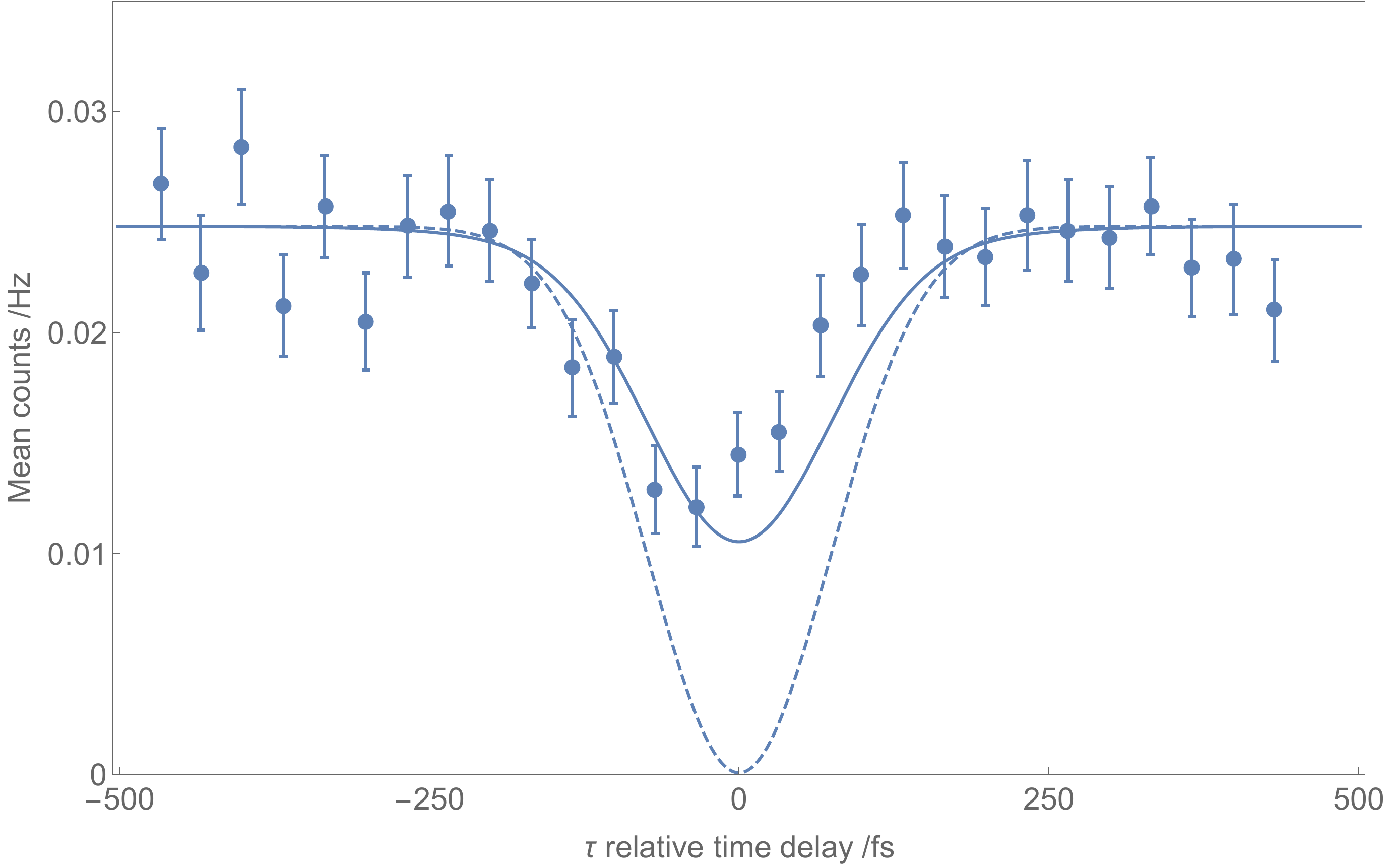}
  \caption{Plot of mean counts for the outputs corresponding to $P_{201}$ as illustrated in Figure~\ref{fig:pseudonumberres}a. The solid line is the fit using our simulation, whilst the dashed line is an ideal theory curve.}
  \label{fig:201aligned}
\end{figure}
\begin{figure}[h!]
\centering
\includegraphics[width=0.8\textwidth]{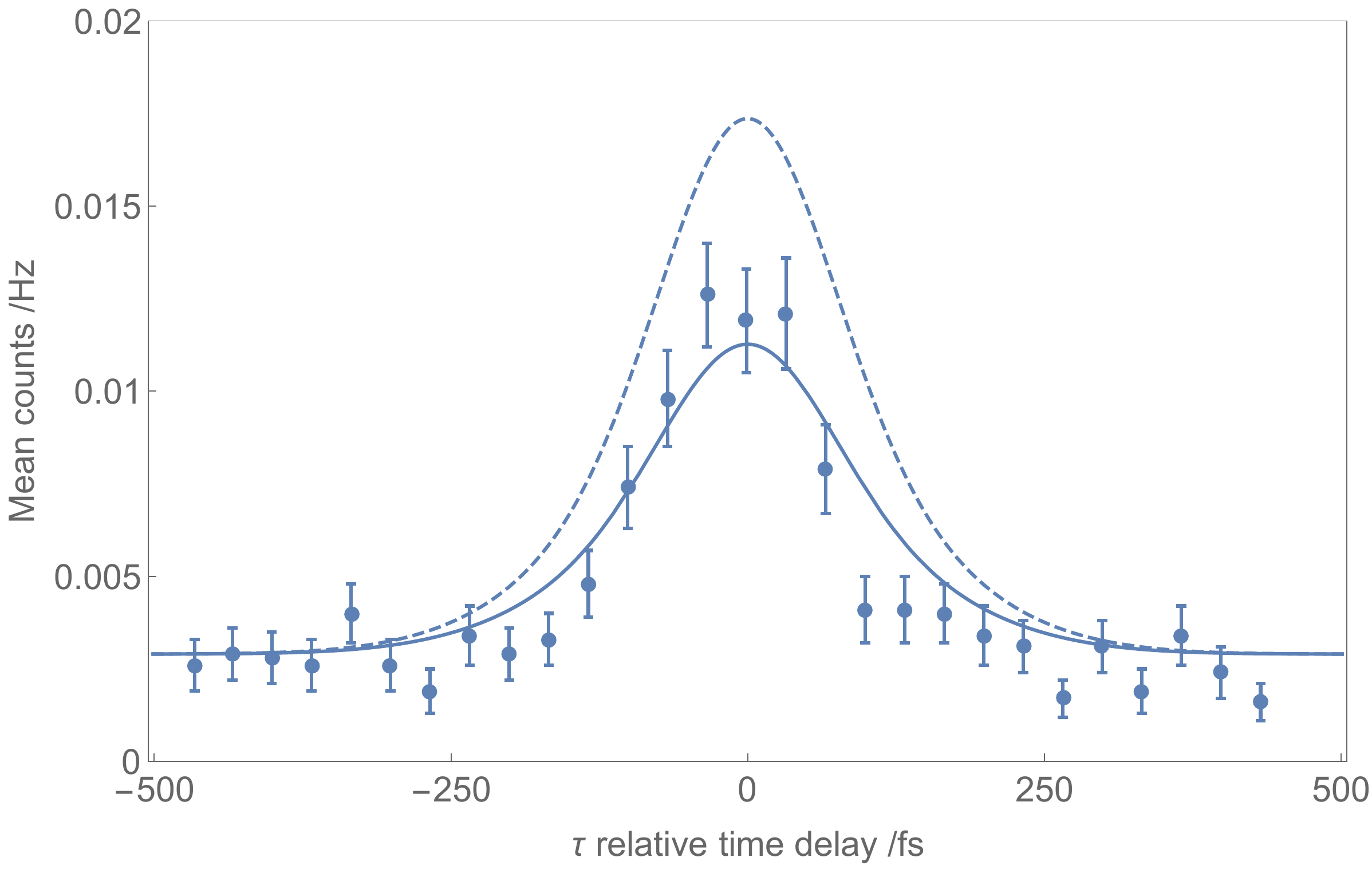}
 \caption{Plot of count rates corresponding to $P_{300}$ as measured using the setup in Figure~\ref{fig:pseudonumberres}a, when all photons have the same polarisation. The solid line is the fit using our simulation and the dashed is an ideal theory curve.}
\end{figure}
\clearpage

\subsection{Polarisations set for $\varphi=\pi$ (cf. Equation 8 in main paper)}
\subsubsection{HOM dips for temporal alignment of photons}
\begin{figure}[h!]
  \centering
  \includegraphics[width=0.63\linewidth]{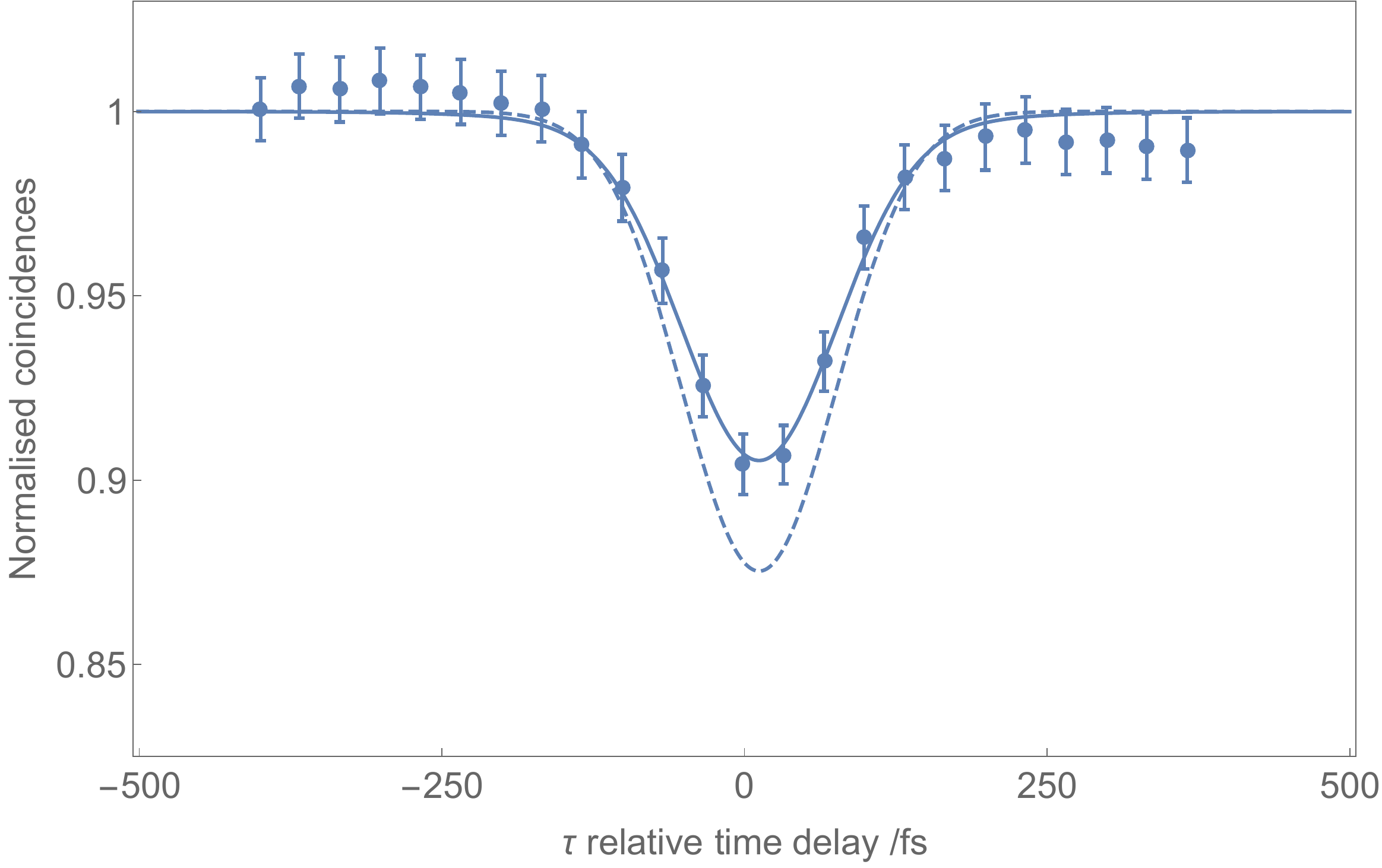}
  \caption{Plot of normalised heralded two-photon coincidences through the tritter when the injected photons have polarisations as in Eqn. 8 of the main paper. In this case we inject photons into the first and second tritter inputs and monitor the first and second output ports. The solid line is the fit using our simulation, whilst the dashed line is an ideal theory curve.}
  \label{fig:HOM1212merc}
\end{figure}

Again to align the three photons temporally before injection into the tritter, we perform HOM measurements for the three pairs of photons. We expect 12.5\% visibility but record closer to 10\%, again due to the effects mentioned in the main paper. The dip in Figure~\ref{fig:HOM1212merc} is twice as narrow as the others, corresponding to the dip between two photons which are both being translated in time on injection. The other two dips are from when only one of the photons injected into the tritter is translated in time (see Figure 3 in main text). The dips are all centred such that the three photons overlap in time when the stages are at their zero positions.

\begin{figure}[h!]
  \centering
  \includegraphics[width=0.63\linewidth]{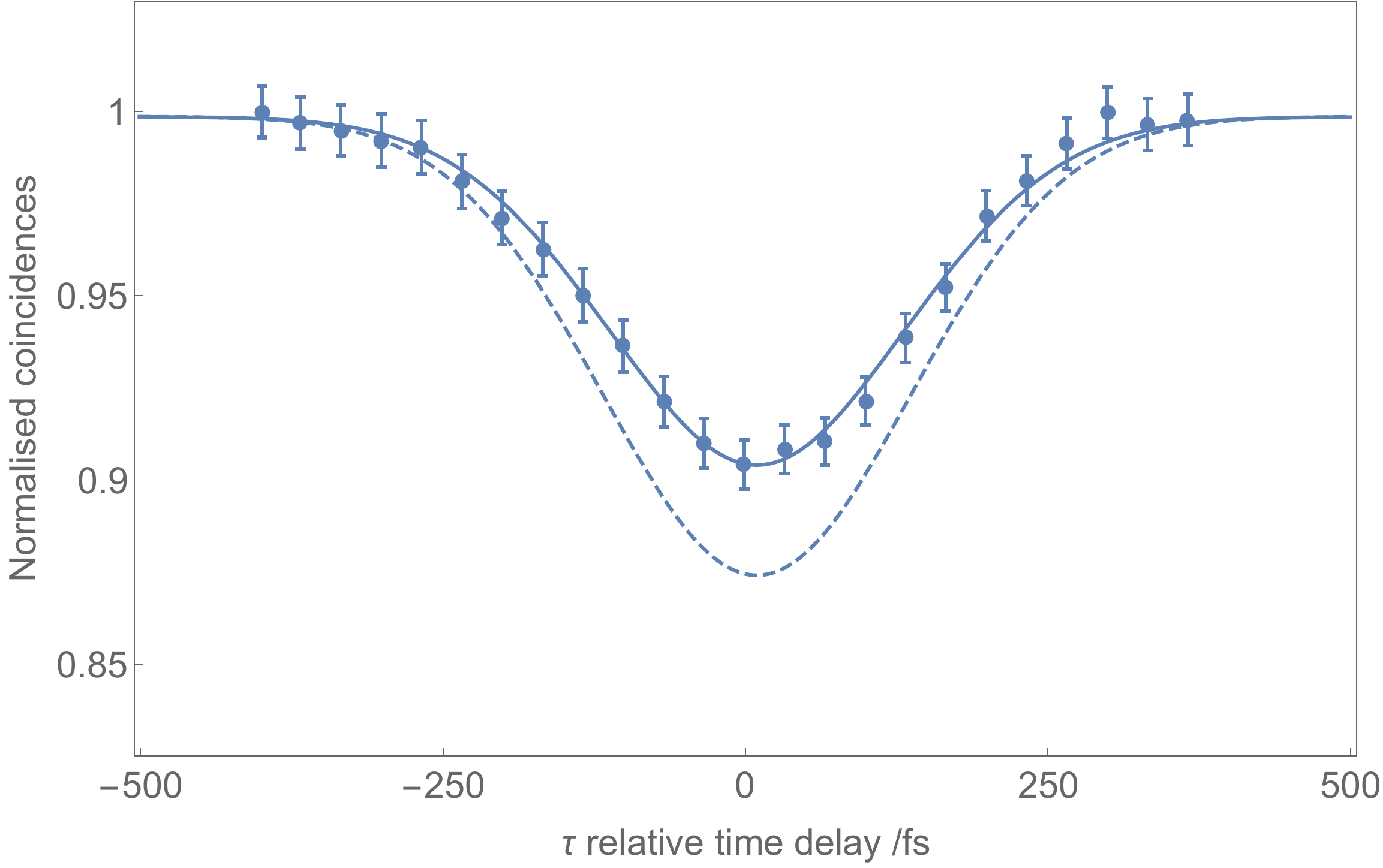}
  \caption{Plot of normalised heralded two-photon coincidences through the tritter when the injected photons have polarisations as in Eqn. 8 of the main paper. In this case we inject photons into the first and third tritter inputs and monitor the first and third output ports. The solid line is the fit using our simulation, whilst the dashed line is an ideal theory curve.}
  \label{fig:HOM1313merc}
\end{figure}
\begin{figure}[h!]
  \centering
  \includegraphics[width=0.63\linewidth]{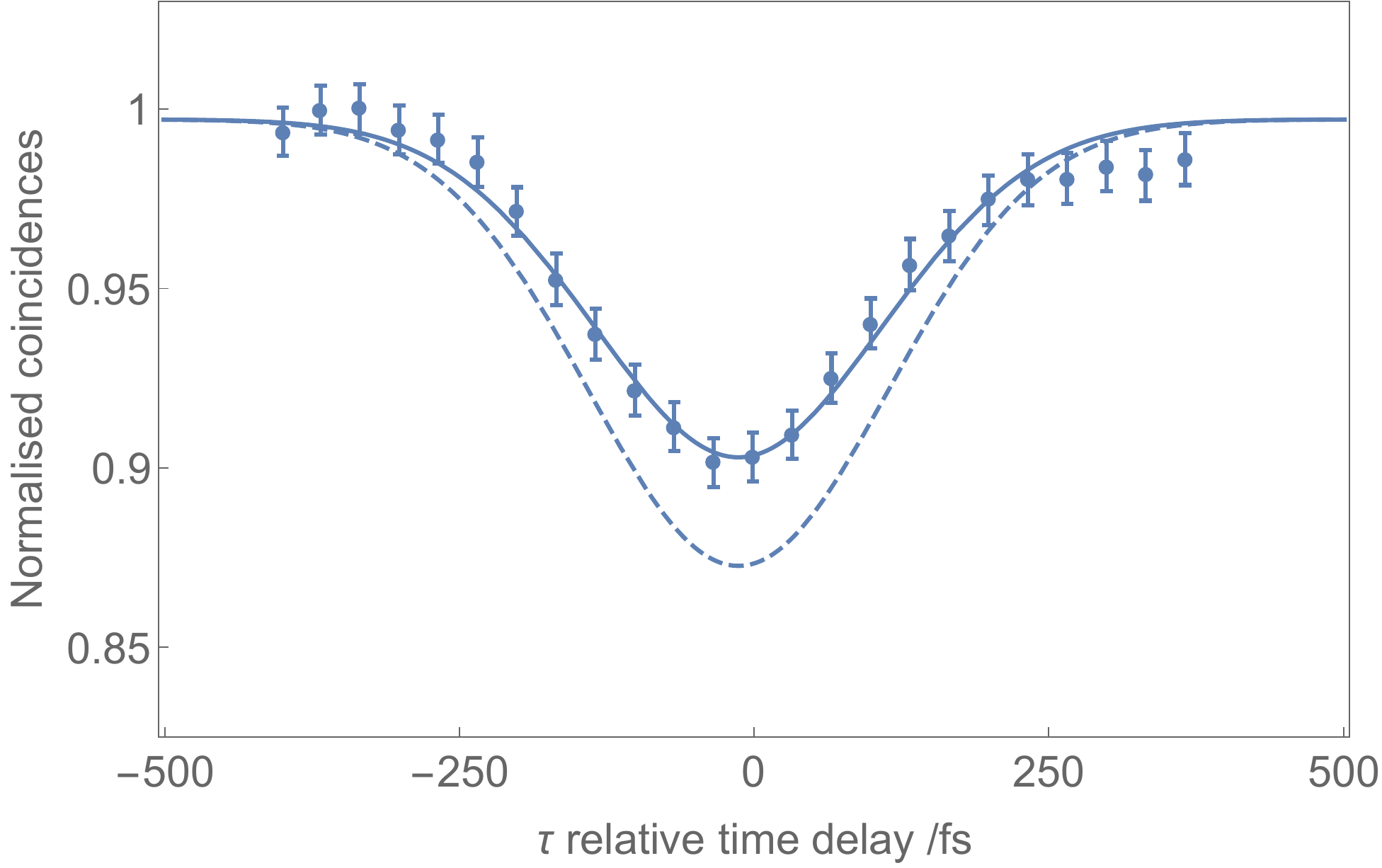}
  \caption{Plot of normalised heralded two-photon coincidences through the tritter when the injected photons have polarisation as in Eqn. 8 of the main paper. In this case we inject photons into the second and third tritter inputs and monitor the second and third output ports. The solid line is the fit using our simulation, whilst the dashed line is an ideal theory curve.}
  \label{fig:HOM2323merc}
\end{figure}
We also recorded coincidences for the outputs corresponding to $P_{210},P_{201},P_{300}$ but these statistics are all predicted to have lower visibilities for this case of $\varphi=\pi$ compared to $\varphi=0$. Our recorded statistics are not sufficient to resolve these features.

\FloatBarrier
\clearpage
\subsection{Probing the triad phase (cf. Equation 9 in main paper)}
\subsubsection{Polarisation dependence of the tritter}
\begin{figure}[h!]
\centering
\includegraphics[width=0.75\textwidth]{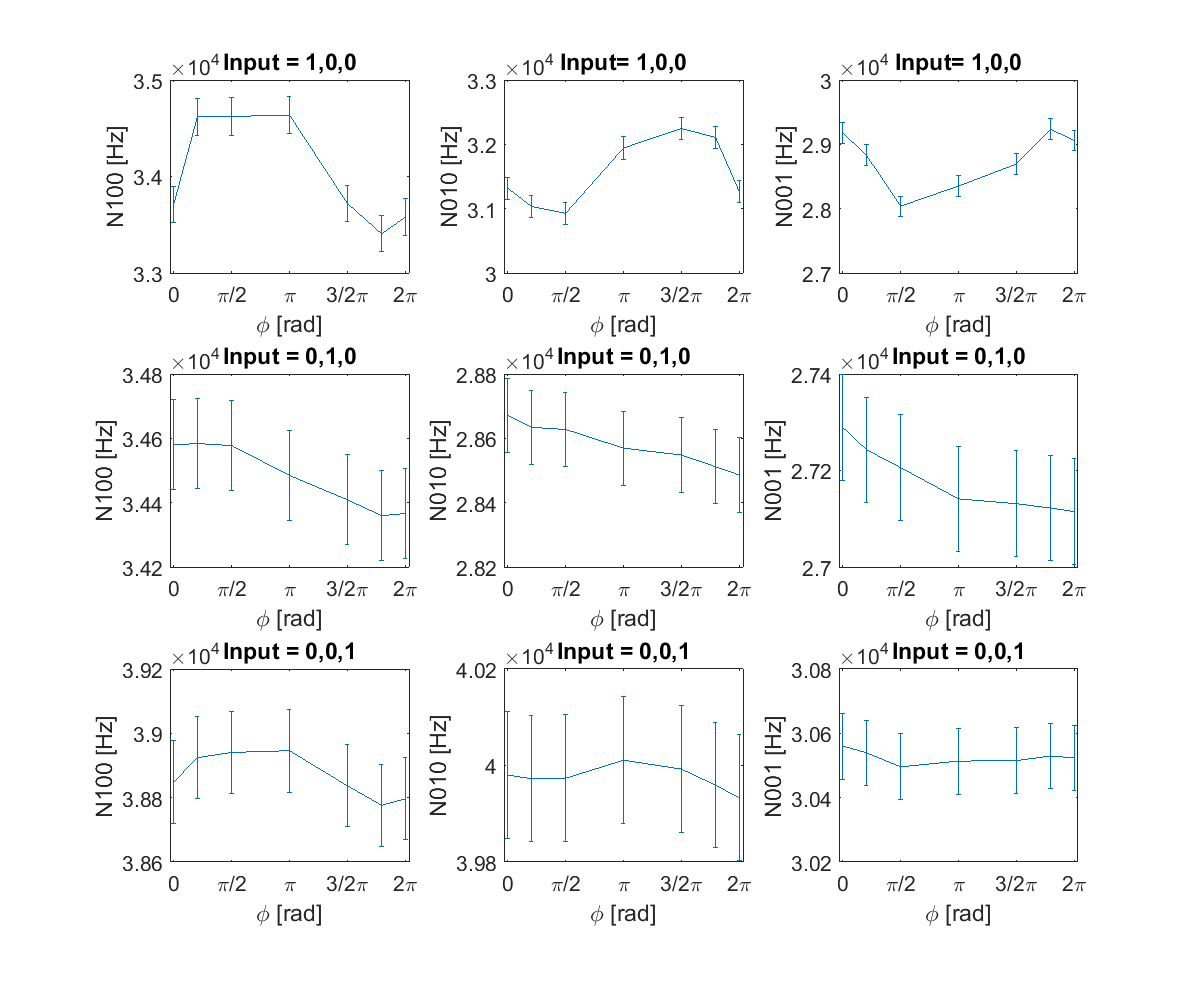}
\caption{\label{fig:SinglesSweep} All input and output combinations for heralded single-photon events. The y-axis labels the count rates for a particular output configuration, and the x-axis is the triad phase we scan. The input port for the injected photon is labelled above each plot. The variation of the counts for the case where the polarisation of the photon is varied before injection (first row) shows that the tritter is slightly polarisation dependent: the coupling between spatial modes varies as a function of the triad phase. The slight drop of counts shown in the second row (where a single photon is injected into the second tritter input) is due to imperfect fibre coupling.}
\end{figure}

\begin{figure}
\centering
\includegraphics[width=0.50\textwidth]{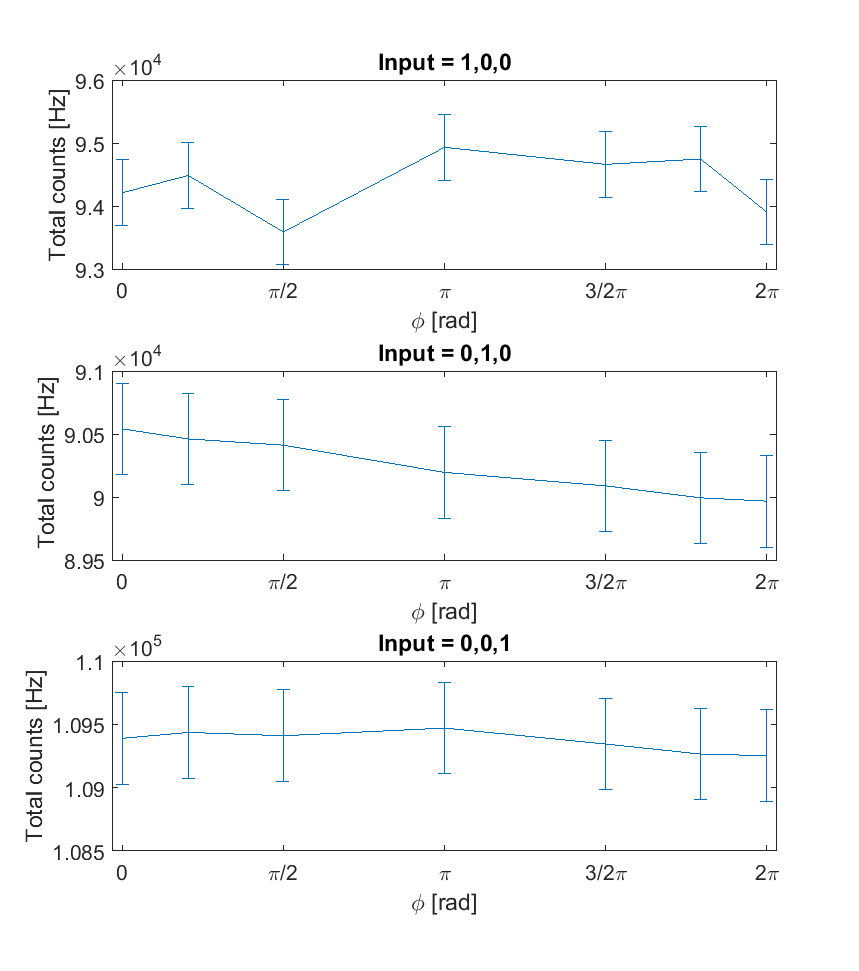}
\caption{\label{fig:SumSinglesSweep} We plot the sum of all heralded single counts for different inputs into the tritter (total counts=N100+N010+N001, corresponding to summing the counts in the rows appearing in Figure~\ref{fig:SinglesSweep}).}
\end{figure}

For isolating three-photon interference, we scan the triad phase by varying the polarisation of one of the photons. In order to study the polarisation-dependence of the tritter, we send heralded single photons into different tritter inputs and record the output counts (see Figures \ref{fig:SinglesSweep} and \ref{fig:SumSinglesSweep}).

The total number of counts is relatively constant (see Fig.~\ref{fig:SumSinglesSweep}), whilst some individual heralded singles events in the bottom row of Figure \ref{fig:SinglesSweep} vary as the triad phase (and thus polarisation of the photon injected into the first input) changes. This suggests that the couplings of the  tritter have a slight polarisation dependence.

\clearpage
\subsubsection{Heralded two-photon coincidences}
\begin{figure}[h!]
\centering
\includegraphics[width=0.75\textwidth]{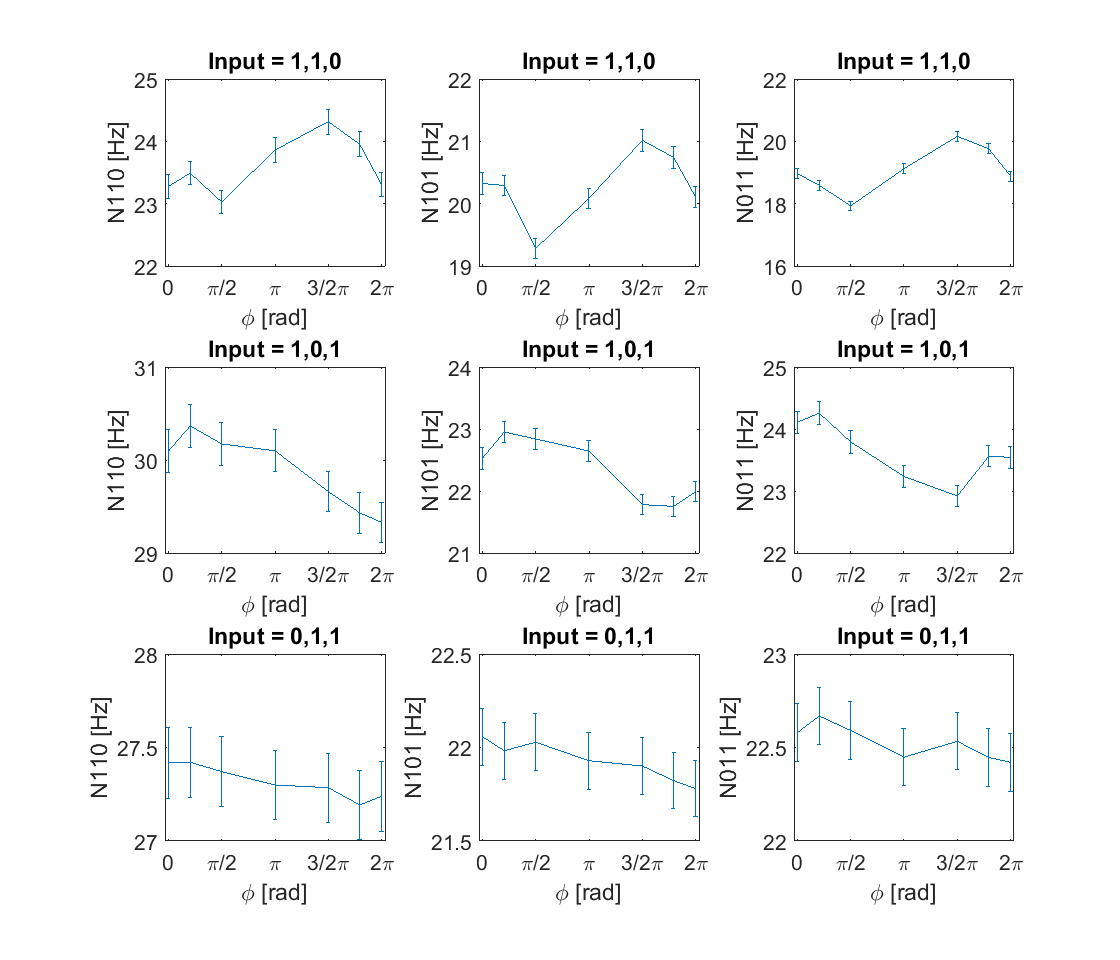}
\caption{\label{fig:HOMDipsSweep} All input and output combinations for heralded two-photon events. We plot the number of heralded two-fold coincidences in the first and second (N110), first and third (N101), and second and third (N011) spatial output modes when changing the triad phase (and thus polarisation of the photon injected into the third input). The channels with the highest variation are those involving the first input channel, and this suggests it is due to the tritter's polarisation dependence.}
\end{figure}

We monitored the heralded two-fold coincidences to verify that we have as little variation as possible as a function of the triad phase. In Figure~\ref{fig:HOMDipsSweep} all possible combinations of heralded two-photon events are displayed. The largest variation in counts is observed for channels containing the first input channel, arising, as discussed in the previous section, from the tritter's polarisation dependence. 

\FloatBarrier
\clearpage
\subsubsection{Additional output event plots}
\begin{figure}[h!]
\centering
\includegraphics[width=1\textwidth]{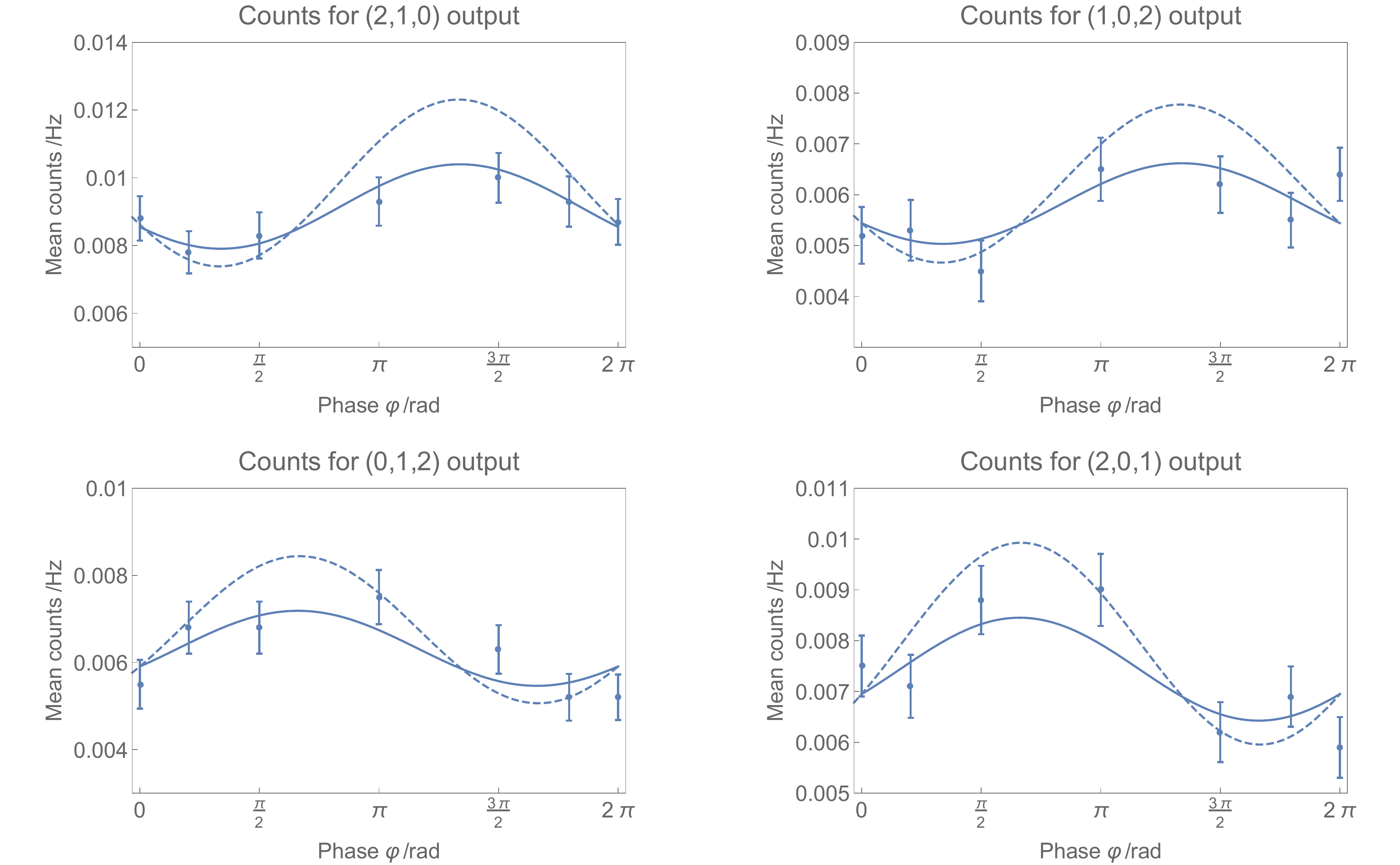}
  \caption{Plots for count rates corresponding to cases where two photons exit the same output port, whilst the third exits in a different port. From Eqn.~\ref{eqn:probs} we expect cosine curves shifted by $-\pi/3$ for (2,1,0) and (1,0,2), and by $+\pi/3$ for (0,1,2) and (2,0,1). The solid lines are simulation curves and the dashed lines are ideal theory, and both have been normalised to fit the data at $\varphi=0,2\pi$ for comparison.} 
\end{figure}
\clearpage
\FloatBarrier

%
\clearpage
\section{Simulation of the experiment}
%
In order to provide a simulation of the experiment, we used the formalism developed in \cite{Tichy2015,Shchesnovich2015,Shchesnovich2015a} to simulate general mixed, squeezed states, contaminated with distinguishable noise photons, that are input into a lossy unitary.

\subsection{Impure input states}

It was noted previously in~\cite{Tichy2015} that the counting statistics for a mixed state input can be expressed as a function of the density matrices $\rho_i$ for each photon in input mode $i$. 
For three photons input to an interferometer described by the unitary $U$ this leads to the following expression for the coincidence probability $P_{111}$:

\begin{equation}
\begin{split}
\label{eqn:P111MixMalte}
P_{111}=\text{perm}(U*U^\star)
+\text{Tr}(\rho_1\rho_2)\text{perm}(U*U_{2,1,3}^\star)
+\text{Tr}(\rho_1\rho_3)\text{perm}(U*U_{3,2,1}^\star)
+\text{Tr}(\rho_2\rho_3)\text{perm}(U*U_{1,3,2}^\star)\\
+2\text{Re}( Tr(\rho_1\rho_2\rho_3))\text{Re}(\text{perm}(U*U_{2,3,1}^\star))
-2\text{Im}( \text{Tr}(\rho_1\rho_2\rho_3))\text{Im}(\text{perm}(U*U_{2,3,1}^\star))
\end{split}
\end{equation}

For simplicity in the simulation we make the assumption that we can decompose the density matrix into a mixed and a pure subspace, where the full density matrix for each photon is given by their tensor product:
\begin{equation}
\label{eqn:model}
\rho_i=\rho_{pure,i}\otimes\rho_{mixed,i}
\end{equation}

$\rho_{pure}$ may be represented as the tensor product of a density matrix which contains the temporal modes and another containing the polarisation degree of freedom. 
\begin{equation}
\label{eqn:model2}
\rho_{pure,i}=\rho_{temp,i}\otimes\rho_{pol,i}
\end{equation}

For general temporal modes $\ket{t_1},\ket{t_2},\ket{t_3}$, we find a representation of the states in terms of orthonormal modes $\ket{\tau_1},\ket{\tau_2},\ket{\tau_3}$ using the Gram-Schmidt decomposition:

\begin{eqnarray}
\ket{t_1}&=&\ket{\tau_1}\\
\ket{t_2}&=&\braket{t_1}{t_2}\ket{\tau_1}+\sqrt{1-|\braket{t_1}{t_2}|^2}\ket{\tau_2}\\
\ket{t_3}&=&\braket{t_1}{t_3}\ket{\tau_1}+\alpha\ket{\tau_2}+\sqrt{1-|\alpha|^2-|\braket{t_1}{t_3}|^2}\ket{\tau_3}
\end{eqnarray}

Where $\alpha=\frac{\braket{t_2}{t_3}-\braket{t_2}{t_1}\braket{t_1}{t_3}}{\sqrt{1-|\braket{t_1}{t_2}|^2}}$

and  $\ket{t_1} ,\ket{\tau_2},\ket{\tau_3}$ are a set of orthonormal vectors. We can then construct the density matrices in mode basis:
\begin{equation}
\rho_{temp,i}=\ket{t_i}\bra{t_i}
\end{equation}

The polarisation density matrix is constructed from basis states $\ket{H}$ and $\ket{V}$. Mixedness is modelled on a two dimensional Hilbert-space which is chosen to be orthogonal to time-frequency and polarisation modes.

\subsection{Higher order photon contributions}

The state of a single ideal two-mode-squeezer is given by:

\begin{equation}
\ket{\Psi}=\sqrt{1-\lambda^2}\sum^\infty_{n=0}\lambda^n\ket{n_{s}n_{i}}
\label{eqn_squeeze1}
\end{equation}

Furthermore, we assume that in each source uncorrelated photons are created with probabilities $P_I$ for the idlers and $P_S$ for the signals. In particular $(1-P_I)(1-P_S)$ is the probability of producing no uncorrelated noise photons.  $(1-P_I)P_I(1-P_S)P_S$ is the probability of creating exactly one uncorrelated photon pair.

We can then construct the density matrix for one source's emission:

\begin{equation}
\hat{\rho}=(1-\lambda^2)\cdot(1-P_I)\cdot(1-P_S)\sum^\infty_{n,k,l=0}\lambda^{2n}P_I^{k}P_S^{l}\ket{n_{s}n_{i},k_{s}l_{i}}\bra{n_{s}n_{i},k_{s}l_{i}}
\label{eqn_squeeze3}
\end{equation}
where for each total number of photons $2n+k+l$, we include cases where they come from four-wave mixing or noise processes. The indices $k$ and $l$ label the number of signal and idler noise photons which are assumed to be completely distinguishable from all other photons.

\subsection{Parameter values}

In the following table we give the parameter values that were used for the simulation:

\begin{center}
\begin{tabular}{ |l |l|l| }
\hline
Name & Symbol & Value \\
\hline
Squeezing-parameter & $\lambda$ & 0.16 \\
 \hline
 Purity & $\mathcal{P}$ & 0.9 \\
\hline
Fluorescence probability idler & $P_I$ & 0.035   \\
  \hline
Fluorescence probability signal & $P_S$ & 0.009   \\
  \hline   
\end{tabular}
\end{center}

The squeezing parameter was taken to be the same as in~\cite{Spring2016}; the experiment reported in~\cite{Spring2016} was performed with the same power of the pump beam). The purity is a lower bound estimate and primarily affected by our ability to filter out non-factorable components in the (signal/idler) joint spectral distribution. We were limited in the signal/idler filtering bandwidth as we used a single pair of angle tuned bandpass filters in the beam path of signal and idler photons, immediately after a dichroic mirror. Since the three beams pass through the filters at slightly different angles the filters' spectral edges are slightly shifted with respect to each other, effectively limiting our tuning range. We calculate the degree of spectral purity for the given filter bandwidth of $10-15$ nm and obtain a value of approximately $\approx 90\%$ purity. The uncorrelated noise probability is obtained from a measurement of the heralded $g^{(2)}(0)$ in \cite{Spring2016} (supplementary). We perform a fit of the $g^{(2)}(0)$ to our model and use $P_I$ as a free parameter. $P_S$ is chosen to be $1/4$ of $P_I$ as the background noise for the signals is significantly smaller. The ratio of $\frac{P_S}{P_I}\approx 0.25$ was obtained by comparing background noise levels of signal and idler photons with a single photon spectrometer. When the pump polarisation is rotated by 90 degree we lose phase-matching, allowing us to observe the background noise only at the given input power.

\end{document}